\documentclass[dvipdfm,a4paper,12pt]{article}

\makeatletter

\@addtoreset{equation}{section}
\makeatletter

\usepackage{amsmath,amssymb,amsthm}
\usepackage{natbib}
\usepackage{bm}
\usepackage{tabularx}
\usepackage{longtable}

\theoremstyle{plain}
\newtheorem{theorem}{Theorem}[section]
\newtheorem{lemma}{Lemma}[section]

\newtheorem{corollary}{Corollary}[section]
\theoremstyle{definition}

\newtheorem{remark}{Remark}[section]

\DeclareMathOperator{\Var}{Var}
\DeclareMathOperator{\conv}{conv}
\DeclareMathOperator{\diag}{diag}
\DeclareMathOperator{\tr}{tr}
\DeclareMathOperator{\const}{const.}
\newcommand{\indep}{\mathop{\perp\!\!\!\!\perp}}

\title{Group Lasso for high dimensional sparse quantile regression models}
\author{Kengo Kato \footnote{
Department of Mathematics, Graduate School of Science, Hiroshima University, 1-3-1 Kagamiyama, Higashi-Hiroshima, Hiroshima 739-8526, Japan.
Email: \texttt{kkato@hiroshima-u.ac.jp}}}

\begin{document}
\maketitle

\begin{abstract}
This paper studies the statistical properties of the group Lasso estimator for high dimensional sparse quantile regression models where the number of explanatory variables (or the number of groups of explanatory variables) is possibly much larger than the sample size while the number of variables in ``active'' groups is sufficiently small. 
We establish a non-asymptotic bound on the $\ell_{2}$-estimation error of the 
estimator. This bound explains  situations under which the group Lasso estimator is potentially superior/inferior to the $\ell_{1}$-penalized quantile regression estimator in terms of the estimation error. 
We also propose a data-dependent choice of the tuning parameter to make the method more practical, by extending the original proposal of \cite{BC10} for the $\ell_{1}$-penalized quantile regression estimator.
As an application, we analyze high dimensional additive quantile regression models. We show that under a set of suitable regularity conditions, the group Lasso estimator can attain the convergence rate arbitrarily close to the oracle rate.
Finally, we conduct simulations experiments to examine our theoretical results. 

\vspace*{1mm}

\noindent {\em AMS2010 subject classifications}: 62G05, 62J99

\end{abstract}

\section{Introduction}

During the last decade, a great deal of attention has been paid for  penalization methods to the estimation of high dimensional sparse statistical models where the number of explanatory variables is possibly larger  than the sample size while the number of ``active'' variables is sufficiently small. The most popular penalization method would be the $\ell_{1}$-penalization which is coined as ``Lasso'' \citep{Ti96} for the linear regression case. A number of researchers have studied the statistical properties such as the $\ell_{2}$-estimation error and the model selection property of the $\ell_{1}$-penalized estimator for high dimensional linear regression models \citep{BTW07a,BTW07b,ZY07, ZH08,W09,MY09,BRT09}, for high dimensional generalized linear models such as logistic regression models \citep{vdG08,NRWY10}, and for high dimensional quantile regression models \citep{BC10}. Another important penalization method is the group Lasso \citep{YL06}, which intends to select groups of  variables instead of selecting  variables individually. 
In this paper, we study the statistical properties of the group Lasso for high dimensional sparse quantile regression models.

Since the seminal work of \cite{KB78}, quantile regression is one of the main topics in statistics and econometrics. An attractive feature of quantile regression is that it allows us to make inference on the entire conditional distribution by estimating 
several different conditional quantiles. We refer to \cite{K05} for a standard textbook on quantile regression. 
The recent work of \cite{BC10} established bounds on the estimation error and the number of selected variables of the $\ell_{1}$-penalized quantile regression estimator. In particular, they established that, with a suitable choice of the tuning parameter, the $\ell_{1}$-penalized quantile regression estimator attains the near oracle 
convergence rate.  Furthermore, they proposed a data-dependent choice of the tuning parameter to make the method more practical. Their work is thought be a breakthrough to the study of penalization methods for high dimensional sparse quantile regression models. 

The contributions of this papers are threefold. The first and main contribution is to establish a non-asymptotic bound on the $\ell_{2}$-estimation error of the group Lasso estimator for high dimensional sparse quantile regression models.
In particular, we derive a bound that can explain situations under which the group Lasso estimator is potentially superior/inferior to the $\ell_{1}$-penalized estimator. 
The group Lasso estimation that we study requires a prior knowledge on the sparsity pattern of the parameter vector, i.e., a prior knowledge that the parameter vector is groupwise sparse. Intuitively, the group Lasso should have a superior estimation performance to the $\ell_{1}$-penalization when the prior knowledge is ``accurate''. Our result formally gives a theoretical support on this intuition. It should be noted that in contrast to \cite{BC10} who focused on the zero bias case where the conditional quantile function has an exact sparse representation of basis functions, we allow for the non-zero bias case where the conditional quantile function may not have an exact sparse representation but can be reasonably well approximated by a sparse linear combination of basis functions. The second contribution, which is less original, is to extend a data-dependent choice of the tuning parameter, originally proposed by \cite{BC10},  to the group Lasso case. 
Although their original proposal is restricted to the zero bias case, we show that the proposed data-dependent choice is asymptotically valid even for the non-zero bias case under suitable conditions. Third, we apply our general results to the estimation of high dimensional sparse additive  quantile regression models. We allow for the possibility that the number of explanatory variables is much larger than the sample size but assume that the number of active variables is fixed.  The additive components are approximated by truncated series expansions with suitable basis functions. With this approximation, the variables selection becomes the group selection of coefficients in this expansion. In this regard, the group Lasso is suited to the estimation of high dimensional additive models. 
We show that under a set of suitable regularity conditions, the group Lasso estimator can attain the convergence rate arbitrarily close to Stone's (1982, 1985) oracle rate $n^{-\nu/(2\nu +1)}$, where $\nu$ indicates the smoothness of the conditional quantile function. Such a result is new in the quantile regression literature. We also conduct simulation experiments to examine our theoretical results. The focus of this paper is on the estimation performance of the group Lasso estimator for quantile regression, and we do not formally discuss its model selection property. 

From a technical point of view, deriving a non-asymptotic bound that can explain the benefit of the group Lasso for the quantile regression case is a delicate issue. 
One technical difficulty is that the objective function of quantile regression is non-differentiable, which implies that some techniques used in the analysis of linear regression models are not directly applicable to the quantile regression case. Furthermore, a naive extension of Bellini and Chernozhukov's proof strategy will lead to a cruder bound that can not explain the benefit of the group Lasso. To that end, we make use of a Bernstein type inequality for vector-valued Rademacher processes, which turns out to be a key technical device to our result. 
We also use some materials on empirical processes theory (such as Talagrand's (1996) concentration inequality) and geometric functional analysis to control the asymptotic behavior of many (possibly) large matrices arising from the group Lasso formulation.

There are a number of papers on the group Lasso.
\cite{B08}, \cite{NR08}, \cite{HZ10}, \cite{WZ10}, \cite{OWJ10} \cite{LPTV10} studied the statistical properties of the group Lasso estimator for linear regression models. \cite{LPTV10} listed some applications in which the group Lasso is potentially useful. 
\cite{MVB08} applied the group Lasso to logistic regression and established the convergence rate of the estimator; however, they did not demonstrate the benefit of the 
group Lasso over the $\ell_{1}$-penalty. 
\cite{NRWY10} established a deterministic bound on the general M-estimator with a decomposable penalty, 
which includes the group Lasso penalty as a special case; however they  focused on smooth objective functions and did not cover the quantile regression case.
They also did not demonstrate the benefit of the group Lasso over the $\ell_{1}$-penalty except for the linear regression case.
It should be noted that at least technically, mean and quantile regressions are significantly different, so that the analysis of the group Lasso for quantile regression requires a separate treatment. 

The application of the group Lasso (and its variants) to the estimation of nonparametric additive models has recently gained a lot of attention \citep{RLLW09,MVB09,HHW10,KY10, RWY10}.
However, all these papers focused on smooth objective functions and did not cover the quantile regression case. \cite{HHW10} derived the convergence rate of the group Lasso estimator for high dimensional sparse additive mean regression models. 
Their rate is $n^{-\nu/(2\nu + 1)} \sqrt{\log (d \vee n)}$ ($d$ is the number of explanatory variables), which may be significantly slower than the optimal rate $n^{-\nu/(2\nu+1)}$ as $d$ may be of an exponential order. Our result improves upon their rate result in the case of quantile regression (it should be noted that, however, the main concern of their paper is not on the standard group Lasso estimator but the adaptive group Lasso estimator) . 

The remainder of the paper is as follows. Section 2 introduces the model, the estimation method, and the computational method of the group Lasso estimate. Section 3 presents the main results. We establish a non-asymptotic bound on the $\ell_{2}$-estimation error of the group Lasso estimator.
Using this  bound, we derive asymptotic bounds on the estimation error in typical situations. We  make a brief comparison of the theoretical performance of the group Lasso and the $\ell_{1}$-penalized estimators. 
We then propose a data-dependent choice of the tuning parameter.
Section 4 contains an application of our general results to the estimation of high dimensional sparse additive quantile regression models. 
Section 5 presents simulation results.

We explain the notation used in the paper. 
For two sequences $a=a(n)$ and $b=b(n)$, we use the notation $a \lesssim b$ 
if there exists a positive constant $C$ independent of $n$ such that $a \leq C b$, $a \asymp b$ if $a \lesssim b$ and $b \lesssim a$, and $a \lesssim_{p} b$ if $a \leq C b$ with probability approaching one.
For $a,b \in \mathbb{R}$, $a \wedge b := \min \{ a,b \}$ and $a \vee b := \max \{ a,b \}$. Let $\mathbb{S}^{d-1}$ denote the unit sphere on $\mathbb{R}^{d}$ for a positive integer $d$. 
Let $\bm{0}_{d}$ and $\bm{1}_{d}$ denote the $d$-dimensional vectors consisting of zeros and ones only, respectively; 
let $\bm{I}_{d}$ denote the $d \times d$ identity matrix. We use $\| \cdot \|_{2}$ to indicate the Euclidean norm, and use $\| \cdot \|_{0}$ to indicate the $\ell_{0}$-seminorm. For a matrix $\bm{A}$, let $\| \bm{A} \|$ denote the operator norm of $\bm{A}$. For a symmetric positive semidefinite matrix $\bm{A}$, let $\bm{A}^{1/2}$ denote the symmetric square root matrix of $\bm{A}$. We sometimes use the notation $\bm{x}_{-1} = (x_{2},\dots,x_{p})'$ for $\bm{x} \in \mathbb{R}^{p}$. 

\section{Preliminaries}

\subsection{Model and estimation method}

We consider the quantile regression model 
\begin{equation}
y_{i} = g(\bm{z}_{i}) + u_{i}, \ \mathrm{P}(u_{i} \leq 0 \mid \bm{z}_{i}) =  \tau, \label{eq:model}
\end{equation}
where $y_{i}$ is a dependent variable, $\bm{z}_{i}$ is a vector of $d$ explanatory variables and $\tau \in (0,1)$ is a quantile index.
We assume that $\tau$ is fixed. Let $\mathcal{Z}$ denote the support of $\bm{z}_{1}$. 

Suppose that we have a set of basis functions $\{ \psi_{j} : j=1,\dots,p \}$ on $\mathcal{Z}$ where $\psi_{1}(\bm{z}) \equiv 1$.
For $\bm{\beta} = (\beta_{1},\dots,\beta_{p})' \in \mathbb{R}^{p}$, we have a series approximation: $g(\bm{z}) \approx \sum_{j=1}^{p} \beta_{j} \psi_{j}(\bm{z}) =: g_{\bm{\beta}}(\bm{z})$. Take a sparse vector $\bar{\bm{\beta}} \in \mathbb{R}^{p}$ such that  its approximation error 
\begin{equation*}
a_{\bar{\bm{\beta}}} := \sup_{\bm{z} \in \mathcal{Z}} | g(\bm{z}) - g_{\bar{\bm{\beta}}}(\bm{z})|
\end{equation*}
is sufficiently small.  In what follows, we view $\bar{\bm{\beta}}$ as a target value to be estimated. If $g(\bm{z})$ can be represented as a finite linear combination of basis functions, we can take $\bar{\bm{\beta}}$ as the true coefficient vector (in that case, $a_{\bar{\bm{\beta}}} = 0$). However, it should be noted that,  even in that case, $\bar{\bm{\beta}}$ can be different from the true coefficient vector, $\bm{\beta}^*$, say. This happens when $\bm{\beta}^{*}$ itself is not sparse but there is a sparse vector $\bar{\bm{\beta}}$ such that its approximation error is small. Another possibility is, of course, that $g(\bm{z})$ can not be represented as a finite linear combination of basis functions. 
For ease of exposition, we refer to the case that $a_{\bar{\bm{\beta}}} = 0$ as ``zero bias case'' and the case that $a_{\bar{\bm{\beta}}} \neq 0$ as ``non-zero bias case''.
Although $\bar{\bm{\beta}}$ is generally not uniquely determined, the results below hold for any $\bar{\bm{\beta}}$ satisfying the restrictions stated later. 


To define the group Lasso estimator, we prepare some notation. Define $\bm{x}_{i} := (\psi_{1}(\bm{z}_{i}),\dots,\psi_{p}(\bm{z}_{i}))'$. As in the usual regression case, we also call $\bm{x}_{i}$ explanatory variables. Let $\{ G_{1}, \dots, G_{q} \}$ be a partition of $\{ 1,\dots, p \}$ such that $G_{1} = \{ 1 \}$, i.e., $\bigcup_{j=1}^{q} G_{j} = \{ 1,\dots, p \}$ such that $G_{1}=\{ 1 \}, G_{k} \neq \emptyset$ for all $k$ and $G_{k} \cap G_{l} = \emptyset$ for all $ k \neq l$.  Throughout the paper, we assume $q \geq 2$. We view each $G_{k}$ as a ``group'' for the explanatory variables $\bm{x}_{i}$. Let $p_{k}$ denote the cardinality of $G_{k}$, i.e., $p_{k} := | G_{k} |$. For $\bm{\beta} \in \mathbb{R}^{p}$, we write $\bm{\beta}_{G_{k}} = ( \beta_{j}, j \in G_{k}) \in \mathbb{R}^{p_{k}}, S(\bm{\beta}) := \{ 1 \} \cup \{ k \in \{ 2,\dots, q \} : \| \bm{\beta}_{G_{k}} \|_{2} > 0 \}$, and use the notation $p_{S} := \sum_{k \in S} p_{k}$ for $S \subset \{ 1,\dots, q \}$. Define $\bm{X} := [ \bm{x}_{1} \ \cdots \ \bm{x}_{n} ]', \hat{\bm{\Sigma}} := n^{-1} \bm{X}'\bm{X}, \bm{\Sigma} := \mathrm{E}[\bm{x}_{1}\bm{x}_{1}'], \bm{X}_{G_{k}} := [ \bm{x}_{1G_{k}} \cdots \bm{x}_{nG_{k}} ]', \hat{\bm{\Sigma} }_{k}:= n^{-1} \bm{X}_{G_{k}}'\bm{X}_{G_{k}}$ and $\bm{\Sigma}_{k} := \mathrm{E}[ \bm{x}_{1G_{k}} \bm{x}_{1G_{k}}']$. 
Working with this notation, we consider the group Lasso estimator: 
\begin{equation}
\hat{\bm{\beta}} := \arg \min_{\bm{\beta} \in \mathbb{R}^{p}} \left [ \frac{1}{n} \sum_{i=1}^{n} \rho_{\tau}(y_{i} - \bm{x}_{i}'\bm{\beta}) +  \frac{\lambda}{n} \sum_{k=2}^{q}   \sqrt{p_{k}} \| \hat{\bm{\Sigma}}_{k}^{1/2} \bm{\beta}_{G_{k}} \|_{2} \right ], \label{primal1}
\end{equation}
where $\rho_{\tau}(u) := \{ \tau - I(u \leq 0) \} u$ is the check function and $\lambda$ is a nonnegative tuning parameter. 
 It should be noted that the constant term $\beta_{1}$ is not penalized, which is standard in the literature. The existence of the group Lasso estimate is always guaranteed, although it may not be unique.
 By the nature of the group Lasso penalty, $\hat{\bm{\beta}}_{G_{k}} = \bm{0}$ for some $k$. This means that the group Lasso can select groups of variables. 
 
We wish to establish  a non-asymptotic bound on the $\ell_{2}$-estimation error $\| \hat{\bm{\beta}} - \bar{\bm{\beta}} \|_{2}$. The main assumptions we make are: (i) the number of elements in the active groups, $p_{S(\bar{\bm{\beta}})}$, is sufficiently smaller than $n$, and at the same time (ii) the approximation error $a_{\bar{\bm{\beta}}}$ is sufficiently small, which means that the conditional quantile function is reasonably well approximated by a function $g_{\bar{\bm{\beta}}}$ with a groupwise sparse vector $\bar{\bm{\beta}}$ (more precisely, we are assuming the existence of a vector satisfying (i) and (ii), and taking $\bar{\bm{\beta}}$ as such a vector). 

We shall mention that $p_{S(\bar{\bm{\beta}})}$ is always larger than or equal to $\| \bar{\bm{\beta}} \|_{0}$ since $\bar{\bm{\beta}}_{G_{k}}$ for $k \in \bar{S}$ may have zero elements. 
The group Lasso presumes a prior knowledge on the sparsity pattern that $\bar{\bm{\beta}}$ is groupwise sparse. 
 It would make sense to say that the prior knowledge  is accurate if $p_{S(\bar{\bm{\beta}})}$ is close to $\| \bar{\bm{\beta}} \|_{0}$. 
Intuitively, it is expected that the performance of the group Lasso depends on the accuracy of the prior knowledge. 
In fact, our theoretical results discussed below give a support on this intuition. 

A word of notation. For $S \subset \{ 1,\dots, q\}$, we use the notation $S_{-1} := S \backslash \{ 1 \}$ and $S^{c} := \{ 1, \dots, q \} \backslash S$. For notational convenience, let $\bar{S} := S(\bar{\bm{\beta}})$. 

\subsection{Computation} 

This subsection is concerned with the computational aspect of the group Lasso problem (\ref{primal1}). Put $\lambda_{k} := \lambda \sqrt{p_{k}}$. 
We observe that the problem (\ref{primal1}) is formulated as:
\begin{align}
&\min_{\bm{\beta},\bm{v},\bm{\eta}^{+},\bm{\eta}^{-}} \tau \sum_{i=1}^{n} \eta_{i}^{+} + (1-\tau) \sum_{i=1}^{n} \eta_{i}^{-} +\sum_{k=2}^{q} \lambda_{k} v_{k} \label{primal2} \\
&\text{s.t.} \ \bm{\eta}^{+} - \bm{\eta}^{-} = \bm{y} - \bm{X} \bm{\beta}, \notag \\
&\qquad \| \hat{\bm{\Sigma}}_{k}^{1/2} \bm{\beta}_{G_{k}} \|_{2} \leq v_{k}, \ k=2,\dots,q, \notag \\
&\qquad \bm{\eta}^{+} \geq \bm{0}, \bm{\eta}^{-} \geq \bm{0}, \notag
\end{align}
where the inequalities are interpreted coordinatewise. 
The problem of such type is called a second order cone programming (SOCP) problem \citep{LVBL98,AG03}. The dual of (\ref{primal2}) reduces to
\begin{align}
&\max_{\bm{a} \in \mathbb{R}^{n},\bm{b} \in \mathbb{R}^{p}}  \ \bm{y}'\bm{a} \label{dual} \\
&\ \text{s.t.} \ b_{1} = \bm{1}_{n}'\bm{a} = 0, \notag \\
&\qquad \| \bm{b}_{G_{k}} \|_{2} \leq \lambda_{k}, k=2,\dots,q, \notag \\
&\qquad \hat{\bm{\Sigma}}^{1/2}_{k} \bm{b}_{G_{k}} = \bm{X}_{G_{k}}'\bm{a}, k=2,\dots,q, \notag \\
&\qquad \bm{a} \in [ \tau-1,\tau ]^{n}. \notag
\end{align}
It is clear that there exist strictly feasible solutions to the primal (\ref{primal2}) and the dual (\ref{dual}) problems. Therefore, optimal solutions to those problems exist \citep[cf.][Theorem 13]{AG03}. 
In practice, we may efficiently solve the problem (\ref{primal1}) by using primal-dual interior point algorithms. For instance, in MATLAB implementation, we may use the SeDuMi package \citep{S99} to solve SOCP problems.

\section{Main results}

\subsection{Conditions}

We first introduce some basic conditions.

\begin{description}
\item[(C1)] $\{ (y_{i},\bm{z}_{i}')' : i=1,2,\dots \}$ are independently and identically distributed (i.i.d.) where the pair $(y_{1},\bm{z}_{1}')'$ satisfies the model (\ref{eq:model}). 
\item[(C2)] Let $F(u | \bm{z})$ denote the conditional distribution function of $u_{1}$ given $\bm{z}_{1} = \bm{z}$. Assume that $F(u | \bm{z})$ has a continuously differentiable  density $f(u | \bm{z})$ such that there exist positive constants $\varrho, c_{f},C_{f}$ and $L_{f}$ such that $c_{f} \leq f(u | \bm{z}) \leq C_{f}$ on $[-\varrho,\varrho] \times \mathcal{Z}$ and $|f'(u | \bm{z})| \leq L_{f}$ on the support of $(u_{1},\bm{z}_{1}')'$, where $f'(u | \bm{z}) := \partial f(u | \bm{z})/ \partial u$. 
\item[(C3)] $\mathrm{E}[\| \bm{x}_{1} \|_{2}^{3}] < \infty$. $\bm{\Sigma}_{k} := \mathrm{E}[\bm{x}_{1G_{k}}\bm{x}_{1G_{k}}'] = \bm{I}_{p_{k}}$ for all $k=2,\dots,q$. 
\item[(C4)] $a_{\bar{\bm{\beta}}} \leq \varrho$ and $\mathrm{E}[f(0 | \bm{z}_{1}) \{ g(\bm{z}_{1}) - g_{\bar{\bm{\beta}}}(\bm{z}_{1})\}] = 0$. 
\end{description}

Condition (C1) defines the data generating process.
Conditions (C2) is standard in the quantile regression literature. 
Condition (C3) is a moment condition on $\bm{x}_{1}$. Given that $\bm{\Sigma}_{k}$ is positive definite, the normalization $\bm{\Sigma}_{k}=\bm{I}_{p_{k}}$ does not lose any generality since we can always rescale the original parameter so that the explanatory variables satisfy this normalization. In fact, given the original parameter $\bm{\beta}^{0}$ and the explanatory variables $\bm{x}^{0}_{i}$, define the rescaled parameter $\bm{\beta} = \bm{D}^{1/2} \bm{\beta}^{0}$ and the rescaled explanatory variables $\bm{x}_{i} = \bm{D}^{-1/2} \bm{x}_{i}^{0}$ where $\bm{D} := \diag (1,\bm{\Sigma}_{2},\dots,\bm{\Sigma}_{q})$. The convergence rate under the original parametrization follows from the relation: $\| \hat{\bm{\beta}}^{0} - \bar{\bm{\beta}}^{0} \|_{2} \leq \kappa^{-1/2} \| \hat{\bm{\beta}} - \bar{\bm{\beta}} \|_{2}$ where $\kappa$ is the mimimum eigenvalue of $\bm{D}$. 	
Condition (C4) is a technical requirement on $\bar{\bm{\beta}}$. The first part of condition (C4) imposes a preliminary bound on the approximation error $a_{\bar{\bm{\beta}}}$. Under condition (C2), this ensures that $f( g_{\bar{\bm{\beta}}}(\bm{z}) - g(\bm{z}) | \bm{z}) \geq c_{f}$ on the support of $\bm{z}_{1}$. The second part of condition (C4) does not lose any generality since we can modify $\bar{\bm{\beta}}$  such that it satisfies the second part of condition (C4) without changing the sparsity pattern and the order of the approximation error. In fact, we can show the next lemma. 

\begin{lemma}
Assume conditions (C1)-(C3). 
For a given $\bar{\bm{\beta}}^{0} \in \mathbb{R}^{p}$, there exists a vector $\bar{\bm{\beta}} \in \mathbb{R}^{p}$ such that 
$\mathrm{E}[f(0 | \bm{z}_{1}) \{ g(\bm{z}_{1}) - g_{\bar{\bm{\beta}}}(\bm{z}_{1})\}] = 0, \ S(\bar{\bm{\beta}})= S(\bar{\bm{\beta}}^{0})$ and $a_{\bar{\bm{\beta}}} \leq 2 a_{\bar{\bm{\beta}}^{0}}$. 
\end{lemma}

For a proof, it suffices to see that the vector $\bar{\bm{\beta}} \in \mathbb{R}^{p}$ defined by $\bar{\beta}_{1} = \bar{\beta}^{0}_{1} + \mathrm{E}[f(0 | \bm{z}_{1})]^{-1} \mathrm{E}[f(0 | \bm{z}_{1}) \{ g(\bm{z}_{1}) - g_{\bar{\bm{\beta}}^{0}}(\bm{z}_{1})\}]$ and $\bar{\bm{\beta}}_{-1} = \bar{\bm{\beta}}^{0}_{-1}$
satisfies the requirements of Lemma 3.1. 
The second part of condition (C4) is used to separate the effect of the trivially small group $G_{1}$. See the proof of Lemma \ref{lem6}.

Define $\gamma \in (0,1)$ by 
\begin{equation*}
1-\gamma = \mathrm{P} \{ \| \hat{\bm{\Sigma}}_{k}^{1/2} - \bm{I}_{p_{k}} \| \leq 0.5,  2 \leq \forall k \leq q \}.
\end{equation*}
The constant $0.5$ is not important.
We implicitly assume that $\gamma$ is small, which means that with a high probability, $\hat{\bm{\Sigma}}_{k}^{1/2} $ are not too much deviated from their population values. 
We will give primitive sufficient conditions to guarantee that $\gamma \to 0$ as $n \to \infty$ ($q$ and $p_{k}$ may depend on $n$).

In what follows, let $c_{0} > 3$ denote a fixed constant. 
Define 
\begin{equation*}
\mathbb{C} := \mathbb{C}(c_{0},\bar{S}) := \{ \bm{\alpha} \in \mathbb{R}^{p} : \sum_{k \in \bar{S}^{c}} \sqrt{p_{k}}  \| \bm{\alpha}_{G_{k}} \|_{2} \leq  c_{0} \sum_{k \in \bar{S}} \sqrt{p_{k}} \| \bm{\alpha}_{G_{k}} \|_{2} \}.
\end{equation*}
The set $\mathbb{C}$ is a cone, i.e., for any $\bm{\alpha} \in \mathbb{C}$ and $c > 0$, $c \bm{\alpha} \in \mathbb{C}$. It consists of vectors $\bm{\alpha} \in \mathbb{R}^{p}$ such that the coordinates of $\bm{\alpha}$ in the set $\bar{S}$ are dominant. Such cones of dominant coordinates play an important role in the 
analysis of penalization methods for high dimensional statistical models. The present definition of $\mathbb{C}$ comes from the fact that with a suitable choice of $\lambda$, $\hat{\bm{\beta}} - \bar{\bm{\beta}}$ concentrates on $\mathbb{C}$ with a high probability. 

We introduce a geometric condition associated with the cone $\mathbb{C}$.
\begin{description}
\item[(C5)] (Restricted eigenvalue condition). $\phi_{\min} := \phi_{\min}(c_{0},\bar{S}) := \inf_{\bm{\alpha} \in \mathbb{C} \cap \mathbb{S}^{p-1}} \| \bm{\Sigma}^{1/2} \bm{\alpha} \|_{2} > 0$. 
\end{description}
Condition (C5) is adapted from the restricted eigenvalue condition of \cite{BC10}. The restricted eigenvalue condition originates from \cite{BRT09}. The value of $c_{0}$ allowed depends on whether condition (C5) is satisfied. Note that condition (C5) is more stringent if $c_{0}$ is larger. In the case that $\bm{\Sigma}$ is positive definite, $c_{0}$ can be any constant larger than $3$. 
We also define $\phi_{\max} :=\phi_{\max}(c_{0},\bar{S}) := \sup_{\bm{\alpha} \in \mathbb{C} \cap \mathbb{S}^{p-1}} \| \bm{\Sigma}^{1/2} \bm{\alpha} \|_{2}$. 

Following \cite{BC10}, we introduce the restricted nonlinear impact (RNI) coefficient to deal with the nonlinearity of the quantile regression problem: define 
\begin{equation*}
r^{*} :=r^{*}(c_{0},\bar{S}) := \frac{c_{f}}{L_{f}\phi_{\max}} \inf_{\bm{\alpha} \in \mathbb{C} \cap \mathbb{S}^{p-1}} \frac{\mathrm{E}[ (\bm{\alpha}'\bm{x}_{1})^{2}]^{3/2}}{\mathrm{E}[| \bm{\alpha}'\bm{x}_{1} |^{3}]}. 
\end{equation*}
The present definition of  the RNI coefficient is a natural extension of the same concept defined in \cite{BC10} to the group Lasso case. 
Thus, their comment on the RNI coefficient basically applies to the present case. Note that under condition (C5), $r^{*}$ is positive.

A motivation to introduce the RNI coefficient is concerned with the identification power of $\bar{\bm{\beta}}$. In fact, for $\delta \in [r_{*},r^{*})$ where $r_{*}$ is defines as $r_{*}:= 6 C_{f}a_{\bar{\bm{\beta}}}/c_{f} \phi_{\min}$, it is shown that 
$\mathrm{E}[ \rho_{\tau}(y_{1} - \bm{x}_{1}'\bm{\beta}) - \rho_{\tau}(y_{1} - \bm{x}_{1}'\bar{\bm{\beta}})] > c_{f}\phi^{2}_{\min} \| \bm{\beta} - \bar{\bm{\beta}} \|^{2}_{2}/6$ whenever $\bm{\beta} - \bar{\bm{\beta}} \in \mathbb{C}$ and $\| \bm{\beta} - \bar{\bm{\beta}} \|_{2} = \delta$ (see the proof of Lemma \ref{lem4}).

\subsection{Bound on the $\ell_{2}$-estimation error}

We give a non-asymptotic bound on the $\ell_{2}$-estimation error $\| \hat{\bm{\beta}} - \bar{\bm{\beta}} \|_{2}$. We introduce some notation used in the statement of the theorem. 
Let $A_{1},A_{2}$ and $B$ be any positive constants. Recall that $p_{S} := \sum_{k \in S} p_{k}$ for $S \subset \{ 1,\dots, q \}$. Define
\begin{align*}
&p_{\min} := \min_{2 \leq k \leq q} p_{k},\\
&c_{1} := \frac{c_{0}+3}{c_{0}-3}, \\
&c_{2} := 12 \sqrt{2} (c_{0}+1), \\
&\Delta := L_{f} \sqrt{n} a_{\bar{\bm{\beta}}}^{2}/2 \vee C_{f}\sqrt{n/p_{\min}} a_{\bar{\bm{\beta}}}, \\
&\lambda_{A} := (4 \sqrt{2} + \Delta + A_{1}) \sqrt{n} + A_{2} \sqrt{n \log q/p_{\min}}, \\
&\epsilon^{*}_{B} := \frac{c_{2}(1+\sqrt{p_{\bar{S}_{-1}}})}{\sqrt{n}} + 4 c_{2} B \sqrt{\left ( 1+ \frac{p_{\bar{S}_{-1}}}{p_{\min}} \right )\frac{\log q}{n}}.
\end{align*}

\begin{theorem}
\label{thm1}
Work with the same notation as above. Assume conditions (C1)-(C5). Take $\lambda = c_{1} \lambda_{A}$. 
Then, with probability at least $1- 2 e^{-A_{1}^{2}/2} - 16 q^{1-A_{2}^{2}/128} - 64 q^{1-B^{2}}- 5 \gamma$, we have
\begin{equation}
\| \hat{\bm{\beta}} - \bar{\bm{\beta}} \|_{2} \leq \left [ \frac{6}{c_{f} \phi_{\min}^{2}} \left (  \epsilon^{*}_{B}  \vee \sqrt{\frac{8\phi_{\max}^{2} }{n}} + \frac{1.5 \lambda \sqrt{p_{\bar{S}_{-1}}}}{n}   \right ) \right]  \vee \frac{6 C_{f} a_{\bar{\bm{\beta}}}}{c_{f} \phi_{\min}} , 
\label{eq:bound}
\end{equation} 
provided that the last expression is smaller than $r^{*}$. 
\end{theorem}
\begin{remark}
When we say ``with probability at least $1-\theta$'' and $\theta > 1$, then this means ``with probability at least zero'' (in that case the statement is interpreted as a null statement). 
\end{remark}

The last condition restricts the magnitude of $p_{\bar{S}}$. A similar condition appears in \citet[][Theorem 2]{BC10}. It also restricts the magnitude of the approximation error $a_{\bar{\bm{\beta}}}$. 
These restrictions will be clearer in the asymptotic situation discussed below.
The  bound (\ref{eq:bound}) is similar in flavor to the bounds  derived in  \citet[][Theorem 5.1]{HZ10}, \citet[][Theorem 3.1]{LPTV10} and \citet[][Corollary 4]{NRWY10}. 
However, all these results are on Gaussian linear regression models.

The proof of the theorem appears in Appendix A. 
The proof is different from that of \cite{HZ10} as they exploited the specific property of the least squares problem. The approach taken is similar in spirit to that taken by \cite{vdG08} and \cite{BC10}. 
The first step of the proof is to establish that $\hat{\bm{\beta}} - \bar{\bm{\beta}}$ concentrates on the cone $\mathbb{C}$ with a specified probability; the second step is to relate 
the bound on the estimation error to the tail behavior of some empirical process over the cone $\mathbb{C}$; the third step is to estimate the tail probability of the empirical process by using the symmetrization inequality, the comparison theorem and some concentration inequality for Rademacher processes. An important difference from \cite{BC10} is that to obtain a sharper bound we use a Bernstein type inequality 
instead of a Hoeffding type inequality to the estimation of the tail probability of Rademacher processes (these two inequalities make no significant difference for the $\ell_{1}$-penalty case). 
Using a Hoeffding type inequality \citep[such as (4.12) in][]{LT91} in the present problem leads to a cruder bound of the form $\const \times \sqrt{p_{\bar{S}}\log q/n}$, which is not satisfactory to our purpose (recall that $p_{\bar{S}} \geq \| \bar{\bm{\beta}} \|_{0}$ and the convergence rate of the $\ell_{1}$-penalized estimator is $\sqrt{\| \bar{\bm{\beta}} \|_{0} \log p/n}$). Another difference is that we allow for the possibility that $g \neq g_{\bar{\bm{\beta}}}$. Thus, we have to take into account of the approximation error.
A minor but not negligible point is that we have to pay a special care for the treatment of the constant term $\beta_{1}$ to separate the effect of the trivially small group $G_{1}$. Clearly, if $p_{\min}$ is replaced by the minimum over  $1 \leq k \leq q$, then the bound becomes $\const \times \sqrt{p_{\bar{S}} \log q/n}$. Thus, to get a sharp bound, it is essential to separate the effect of the trivially small group $G_{1}$.

To gain a clearer intuition on the bound, it is useful to consider the asymptotic situation.
In what follows, we assume that all parameter values are indexed by the sample size $n$, and take the limit as $n \to \infty$. For instance, $q=q(n)$. We additionally assume:

\begin{description}
\item[(C6)] $1 \lesssim \phi_{\min}$ and $\phi_{\max} \lesssim 1$. 
\item[(C7)] $\gamma \to 0$ as $n \to \infty$.
\item[(C8)] $a^{2}_{\bar{\bm{\beta}}} \lesssim \displaystyle \frac{1}{\sqrt{n}} \wedge \frac{p_{\min}}{n} \wedge \frac{p_{\bar{S}}}{n} \left ( 1+\frac{\log q}{p_{\min}} \right)$. 
\end{description}

Condition (C6) is trivially satisfied if the smallest eigenvalue of $\bm{\Sigma}$ is bounded away from zero uniformly over $n$, and the largest eigenvalue of $\bm{\Sigma}$ is bounded uniformly over $n$. 
Condition (C7) is a high level condition. 
We will give primitive sufficient conditions to ensure condition (C7). It should be noted that condition (C7) implicitly restricts the growth rate of $q$ and $p_{k}$. Condition (C8) restricts the decreasing rate of $a_{\bar{\bm{\beta}}}$. It restricts $a_{\bar{\bm{\beta}}}$ to go to zero sufficiently fast.  Condition (C8) ensures that $\Delta \lesssim 1$ and 
\begin{equation*}
a_{\bar{\bm{\beta}}} \lesssim \sqrt{ \frac{p_{\bar{S}}}{n} \left ( 1+ \frac{\log q}{p_{\min}} \right)}.
\end{equation*} 
For simplicity of exposition,  we here assume that the approximation error is at most of the same order as the estimation error.

\begin{corollary}
\label{cor1}
Assume conditions (C1)-(C8) where the constants $\varrho, c_{f},C_{f}$ and $L_{f}$ are independent of $n$.
\begin{enumerate}
\item[(i)]
Take $\lambda = t \sqrt{n}(1 + \sqrt{\log q/p_{\min}})$ where $t=t(n)$ is an arbitrary sequence such that $t \to \infty$ as $n \to \infty$. Then, as $n \to \infty$, 
\begin{equation*}
\| \hat{\bm{\beta}} - \bar{\bm{\beta}} \|_{2} \lesssim_{p}  t  \sqrt{ \frac{p_{\bar{S}}}{n} \left ( 1+\frac{\log q}{p_{\min}} \right)},
\end{equation*}
provided that the last expression is of order $o(r^{*})$. 
\item[(ii)]
In the case that $q \to \infty$ as $n \to \infty$, take $\lambda = \sqrt{t n} + A\sqrt{n \log q/p_{\min}}$ with a sequence $t = t(n) \to \infty$ and a constant $A > 16 \sqrt{2}$ independent of $n$. Then,  we have 
\begin{equation*}
\| \hat{\bm{\beta}} - \bar{\bm{\beta}} \|_{2} \lesssim_{p}  \sqrt{  \frac{p_{\bar{S}}}{n} \left ( t+\frac{\log q}{p_{\min}} \right)},
\end{equation*}
provided that the last expression  is of order $o(r^{*})$.
\end{enumerate}
\end{corollary}

The corollary is immediate from Theorem \ref{thm1}.
The last condition  restricts the growth rate of $p_{\bar{S}}$. In the optimal case, $1 \lesssim  r^{*}$. For instance, it is true for the case when $\bm{x}_{1,-1}$ has a log concave 
density \citep[][Comment 2.2]{BC10}. In that case, the corollary holds when 
\begin{equation*}
\frac{p_{\bar{S}}}{n} \left ( 1+\frac{\log q}{p_{\min}} \right) \to 0.
\end{equation*}
In the case that $\| \bm{x}_{1G_{k}}/\sqrt{p_{k}} \|_{2}$ is uniformly bounded, i.e., $\| \bm{x}_{1G_{k}}/\sqrt{p_{k}} \|_{2} \leq K$ for some constant $K$ independent of $n$, then, for $\bm{\alpha} \in \mathbb{C}$,
\begin{align*}
| \bm{\alpha}'\bm{x}_{1} | &\leq  K \sum_{k=1}^{q} \sqrt{p_{k}} \| \bm{\alpha}_{G_{k}} \|_{2} \\
&\leq (1+c_{0}) K \sum_{k \in \bar{S}} \sqrt{p_{k}} \| \bm{\alpha}_{G_{k}} \|_{2}  \\
&\lesssim \sqrt{p_{\bar{S}}} \| \bm{\alpha} \|_{2}.
\end{align*}
Thus, $1/\sqrt{p_{\bar{S}}} \lesssim r^{*}$ under condition (C6). In that case, the corollary holds when 
\begin{equation*}
\frac{p^{2}_{\bar{S}}}{n} \left ( 1+\frac{\log q}{p_{\min}} \right) \to 0.
\end{equation*}

When all the groups $G_{2},\dots,G_{q}$ are of equal size, i.e., $p_{2} = \cdots = p_{q}=:m$, then 
\begin{equation*}
\| \hat{\bm{\beta}} - \bar{\bm{\beta}} \|_{2}  \lesssim_{p} t \sqrt{ \frac{p_{\bar{S}}}{n} + \frac{ | \bar{S} | \log q }{n}  }.
\end{equation*}
in the case (i),  and 
\begin{equation*}
\| \hat{\bm{\beta}} - \bar{\bm{\beta}} \|_{2}  \lesssim_{p} \sqrt{ t \frac{p_{\bar{S}}}{n} + \frac{ | \bar{S} | \log q  }{n} }
\end{equation*}
in the case (ii). In this case, we have a plausible interpretation on the bound. The first part, $\sqrt{p_{\bar{S}}/n}$, reflects the difficulty of estimating $p_{\bar{S}}$ (=the number of variables in true active groups) parameters, while the second part, $\sqrt{| \bar{S} | \log q/n}$, reflects the difficulty of finding active groups from total $q$ groups.

The corollary (roughly) recovers the result of \cite{BC10} on the convergence rate of the $\ell_{1}$-penalized quantile regression estimator. In fact, when $p_{k} = 1$ for all $k$, then $q=p$ and $\hat{\bm{\beta}}$ reduces to the $\ell_{1}$-penalized 
quantile regression estimator. Consider the zero bias case, i.e., $a_{\bar{\bm{\beta}}} = 0$.  In that case, the corollary shows that when $p \to \infty$, for $\lambda=A \sqrt{n \log p}$ with a constant $A > 16 \sqrt{2}$ independent of $n$, $\| \hat{\bm{\beta}} - \bar{\bm{\beta}} \|_{2} \lesssim_{p} \sqrt{\| \bar{\bm{\beta}} \|_{0}\log p/n}$, provided that the right side is of order $o(r^{*})$ ($t$ is chosen such that $t \lesssim \log p$). 

The corollary explains situations under which the group Lasso estimator is potentially superior/inferior to the $\ell_{1}$-penalized estimator.
If $p_{\bar{S}}/\| \bar{\bm{\beta}} \|_{0} = o(\log p)$ and $p_{\bar{S}}/p_{\min} = o( \| \bar{\bm{\beta}} \|_{0} )$, which means that the number of ``inactive'' elements in the active groups are relatively small, then the group Lasso estimator has an improved bound over the $\ell_{1}$-penalized estimator. This fits in the intuition that the group Lasso estimator should have a better performance when a prior knowledge on the sparsity pattern is accurate.  Another good news to the group Lasso is that in some cases it can attain the convergence rate arbitrarily close to the oracle rate $\sqrt{p_{\bar{S}}/n}$,  the rate which could be achieved when the true sparsity pattern $\bar{S}$ were known \cite[see][for convergence rates of general M-estimators in presence of diverging number of parameters]{HS00}. On the other hand, if $\| \bar{\bm{\beta}} \|_{0} \log p = o(p_{\bar{S}})$, the group Lasso is possibly inferior. It is also worthwhile to remark that the bound depends on the smallest group (except for the first group). To be precise, 
$p_{\min}$ in the bound can be replaced by the minimum value in a set of ``large'' $p_{k}$s with a suitable change to the choice of $\lambda$. Suppose that some groups, say $G_{2},\dots,G_{q_{1}}$,  have small sizes, i.e., $p_{k} \lesssim 1$ uniformly over $k=2,\dots,q_{1}$. As long as $q_{1} \lesssim 1$, $p_{\min}$ in the bound can be replaced by the minimun value in $p_{q_{1}+1},\dots,p_{q}$.  This modification is straightforward. In the course of the proof of Theorem \ref{thm1}, we just have to separate the effect of $G_{2},\dots,G_{q_{1}}$, as we do for $G_{1}$ in the present proof (see the proofs of Lemmas \ref{lem5} and \ref{lem6}). However, the group Lasso of the present definition may not work well (more precisely, its performance does not exceed that of the $\ell_{1}$-penalized estimator) if there are many small groups in $G_{2},\dots,G_{q}$. These observations are in lines with \cite{HZ10}. 

We end this section with giving primitive sufficient conditions to ensure condition (C7). The proofs of Lemmas \ref{lem1} and \ref{lem2} below appear in Appendix B. 
We shall comment that giving a sharp condition on the growth rate of $q$ and $p_{k}$ is not a trivial task. In fact, sharp conditions for consistency (in the operator norm) of high dimensional sample covariance matrices under various 
distributional conditions are extensively studied in the recent geometric functional analysis literature \citet[see][for review]{V11}. 
We first consider the case that $\| \bm{x}_{1G_{k}}/\sqrt{p_{k}} \|_{2}$ is uniformly bounded. 

\begin{lemma}
\label{lem1}
Assume conditions (C1) and (C3). 
Suppose that there exists a positive constant $K$ independent of $n$ such that $\| \bm{x}_{1G_{k}}/\sqrt{p_{k}} \|_{2} \leq K$ almost surely for all $2 \leq k \leq q$. 
Then, condition (C7) holds if $p_{\max} \log (q \vee n)/n \to 0$, where $p_{\max} := \max_{2 \leq k \leq q} p_{k}$. 
\end{lemma}

The proof of Lemma \ref{lem1} relies on the combination of Talagrand's (1996) concentration inequality for empirical processes and Rudelson's (1999) inequality for Gram matrices. 
A careful examination of the proof gives an exact bound on $\gamma$. 
Given that $\| \bm{x}_{1G_{k}}/\sqrt{p_{k}} \|_{2}$ is uniformly bounded, the growth condition on $q$ and $p_{\max}$ is considerably weak. 
A primitive sufficient condition for the uniform boundedness of $\| \bm{x}_{1G_{k}}/\sqrt{p_{k}} \|_{2}$ is that each element in $\bm{x}_{1}$ is uniformly bounded. 

We next consider the case that $\bm{x}_{1}$ satisfies a subgaussian condition.  For a real-valued random variable $X$, we define 
\begin{equation*}
\| X \|_{\psi_{2}} := \inf \{ s > 0 : \mathrm{E}[\exp (X^{2}/s^{2}) ] \leq 2 \}.
\end{equation*}
We refer to \citet[][Section 2.1]{VW96} for the $\psi_{2}$-norm. Recall that $\bm{\Sigma} := \mathrm{E}[\bm{x}_{1}\bm{x}_{1}']$. 
\begin{description}
\item[(C9)]
$\bm{x}_{i}$s are generated as $\bm{x}_{i} = \bm{\Sigma}^{1/2} \tilde{\bm{x}}_{i}$ where $\tilde{\bm{x}}_{i}$s are i.i.d.  with
$\mathrm{E}[\tilde{\bm{x}}_{1}\tilde{\bm{x}}_{1}']=\bm{I}_{p}$ and $\sup_{\bm{\alpha} \in \mathbb{S}^{p-1}} \| \bm{\alpha}'\tilde{\bm{x}}_{1} \|_{\psi_{2}} \leq C_{\psi}$ for some constant $C_{\psi}$ independent of $n$.  
\end{description}
It is worthwhile to note that under condition (C6), condition (C9)  ensures that $\mathrm{E}[|\bm{\alpha}'\bm{x}_{1}|^{3}] \lesssim \mathrm{E}[(\bm{\alpha}'\bm{x}_{1})^{2}]^{3/2}$, which implies that $1 \lesssim r^{*}$. 
\begin{lemma}
\label{lem2}
Assume conditions (C3) and (C9). Then, condition (C7) holds if $(p_{\max} \vee \log q)/n \to 0$.
\end{lemma}

The proof of Lemma \ref{lem2} relies on a result from geometric functional analysis \citep{MPT-J07}. 
The case that $\bm{x}_{1,-1}$ has a centered normal distribution is an example that satisfies condition (C9) but does not satisfies the condition of Lemma \ref{lem1}.
It should be noted that Lemma \ref{lem2} does not cover Lemma \ref{lem1}. Suppose, for the illustrative purpose, that
$q=2$ and $\bm{x}_{1,-1}$ has a uniform distribution over the finite set $\{ \sqrt{p-1} \bm{e}_{1},\dots, \sqrt{p-1}\bm{e}_{p-1} \}$ where $\{ \bm{e}_{j} \}_{j=1}^{p-1}$ is
the canonical basis on $\mathbb{R}^{p-1}$.  In that case, $\mathrm{E}[\bm{x}_{1,-1} \bm{x}_{1,-1}'] = \bm{I}_{p-1}$ and $\| \bm{x}_{1,-1}/\sqrt{p-1} \|_{2} =1$ but the supremum of the $\psi_{2}$-norm of $\bm{\alpha}'\bm{x}_{1,-1}$ over $\bm{\alpha} \in \mathbb{S}^{p-2}$ diverges as $p \to \infty$.

\subsection{Data-dependent choice of the tuning parameter}

\cite{BC10} proposed a data-dependent choice of the tuning parameter for the $\ell_{1}$-penalized quantile regression estimator. Their proposal can be extended to the group Lasso case. 

By the proof of Theorem \ref{thm1}, it is seen that $\lambda$ should be taken such that the probability of the event $\{ \lambda \geq c_{1} \Lambda \}$ is close to one (see Appendix A for the definition of $\Lambda$). In fact, the constant choice $\lambda = c_{1}\lambda_{A}$ is taken as an upper bound on the $(1-2 e^{-A_{1}^{2}/2} - 16 q^{1-A_{2}^{2}/128})$-quantile of $c_{1} \Lambda$ (see Lemma \ref{lem6}). Thus, a suitable approximation to a high quantile of $c_{1} \Lambda$ will work in place of $\lambda = c_{1} \lambda_{A}$. In this regard, we consider to approximate a high conditional quantile of $\Lambda$ given $\bm{z}_{1}^{n} := \{ \bm{z}_{1},\dots,\bm{z}_{n} \}$. Although $\Lambda$ is unknown and the conditional distribution of $\Lambda$ given $\bm{z}_{1}^{n}$ is not pivotal (i.e., it depends on unknown parameters) in presence of the approximation error $a_{\bar{\bm{\beta}}}$, as long as $a_{\bar{\bm{\beta}}}$ is small, it is expected that $\Lambda$ is close to 
\begin{equation*}
\tilde{\Lambda} : = 
\max_{1 \leq k \leq q} \|  \sum_{i=1}^{n} \{ \tau - I(u_{i} \leq 0) \} (\hat{\bm{\Sigma}}_{k}^{-1/2}\bm{x}_{i G_{k}}/\sqrt{p_{k}}) \|_{2},
\end{equation*}
where $\hat{\bm{\Sigma}}_{k}^{-1/2}$ is interpreted as the generalized inverse of $\hat{\bm{\Sigma}}_{k}^{1/2}$ if it is singular. 
The conditional distribution of $\tilde{\Lambda}$ given $\bm{z}_{1}^{n}$ is pivotal since $I(u_{i} \leq 0)$ are i.i.d. Bernoulli random variables with probability $\tau$ independent of $\bm{z}_{1}^{n}$. 
Let $\tilde{\Lambda} (1-\theta | \bm{z}_{1}^{n})$ denote the conditional $(1-\theta)$-quantile of $\tilde{\Lambda}$ given $\bm{z}_{1}^{n}$.
Since the conditional distribution of $\tilde{\Lambda}$ given $\bm{z}_{1}^{n}$ is pivotal, $\tilde{\Lambda}(1-\theta | \bm{z}_{1}^{n})$ is computable by simulation. 
It thus makes sense to take $\lambda = c_{1} \tilde{\Lambda}(1-\theta | \bm{z}_{1}^{n})$ for small $\theta \in (0,1)$, as it is expected that 
$\mathrm{P}(\lambda \geq c_{1} \Lambda) \approx \mathrm{P}(\lambda \geq c_{1} \tilde{\Lambda}) = 1-\theta$. 

To summarize, we propose to implement the group Lasso as follows. 
\begin{enumerate}
\item Determine a small $\theta \in (0,1)$, e.g., $\theta=0.1$, and a constant $c > 0$.
\item Compute $\tilde{\Lambda}(1-\theta | \bm{z}_{1}^{n})$ by simulation.
\item Compute the group Lasso estimate (\ref{primal1}) for $\lambda = c \tilde{\Lambda}(1-\theta | \bm{z}_{1}^{n})$.
\end{enumerate}
A recommended choice of $c$ is a slightly larger value than $1$, e.g. $c = 1.1$. 
In the remainder of this subsection, we discuss a theoretical validity of the proposed data-dependent choice of the tuning parameter. We separately consider the zero bias and non-zero bias cases. 

{\em Zero bias case}: 
Suppose first that $a_{\bar{\bm{\beta}}} = 0$. In this case, $\tilde{\Lambda} = \Lambda$. Thus, for $\lambda = c_{1} \tilde{\Lambda}(1-\theta | \bm{z}_{1}^{n})$, 
we have $\mathrm{P}(\lambda \geq c_{1} \Lambda \mid \bm{z}_{1}^{n})=1-\theta$. Let $A_{1}$ and $A_{2}$ be any positive constants such that 
$2 e^{-A_{1}^{2}/2} + 16 q^{1-A_{2}^{2}/128} \leq \theta$. Then, by Lemma \ref{lem6}, $\tilde{\Lambda}(1-\theta | \bm{z}_{1}^{n}) \leq \lambda_{A} = (4 \sqrt{2}+A_{1})\sqrt{n} + A_{2} \sqrt{n \log q/p_{\min}}$. Therefore, in view of the proof of Theorem \ref{thm1}, we obtain the next corollary.  

\begin{corollary}
\label{cor2}
Assume conditions (C1)-(C5) with $a_{\bar{\bm{\beta}}} = 0$. For a given $\theta \in (0,1)$, take $\lambda = c_{1} \tilde{\Lambda} (1-\theta | \bm{z}_{1}^{n})$. Let $A_{1}, A_{2}$ and $B$ be any positive constants such that $2 e^{-A_{1}^{2}/2} + 16 q^{1-A_{2}^{2}/128} \leq \theta$. Then, with probability at least $1-64q^{1-B^{2}}-\theta-5\gamma$, the inequality (\ref{eq:bound}) holds with $\lambda$ replaced by $c_{1} \lambda_{A}$, provided that the upper bound is smaller than $r^{*}$. 
\end{corollary}

The results analogous to Corollary \ref{cor1} hold with suitable modifications. Although  the constant choice $\lambda = c_{1} \lambda_{A}$ is available in the zero bias case once $c_{1}$ is determined, we recommend to use the above data-dependent choice because the constant choice $\lambda = c_{1} \lambda_{A}$ is only an upper bound on the (conditional) $(1-2 e^{-A_{1}^{2}/2} - 16 q^{1-A_{2}^{2}/128})$-quantile of $c_{1} \Lambda$, 
and may be too large in practice. This point is discussed in \citet[][Comment 2.1]{BC09} in a different context.

\vspace*{1mm}

{\em Non-zero bias case}: In this case, since $a_{\bar{\bm{\beta}}} \neq 0$, $\tilde{\Lambda}$ is not equal to $\Lambda$. Thus, the conclusion of Corollary \ref{cor2} does not hold. However, as long as $a_{\bar{\bm{\beta}}} \to 0$ sufficiently fast, it is expected that $\lambda \asymp \tilde{\Lambda}(1-\theta | \bm{z}_{1}^{n})$ with a suitable sequence $\theta = \theta (n) \to 0$ gives an asymptotically correct choice. In fact, we can show the next theorem. 

\begin{theorem}
\label{thm3}
Assume condition (C1).
For any sequence $t=t(n) \to \infty$ such that $t n^{-1/2} \sqrt{1+\log q/p_{\min}} \to 0$, take $\theta=(e \vee q^{1/p_{\min}})^{-t^{2}}$. Then, there exists a positive constant $M$ such that for large $n$, with probability one, 
\begin{equation*}
M^{-1} t\sqrt{n}(1+\sqrt{\log q/p_{\min}}) \leq \tilde{\Lambda}(1-\theta | \bm{z}_{1}^{n}) \leq M t\sqrt{n}(1+\sqrt{\log q/p_{\min}}).
\end{equation*}
Therefore, the conclusion of Corollary \ref{cor1} (i) holds for $\lambda =c \tilde{\Lambda}(1-\theta | \bm{z}_{1}^{n})$ where $c$ is any positive constant.
\end{theorem}

The proof of Theorem \ref{thm3} appears in Appendix C. Theorem \ref{thm3} shows that the proposed data-dependent choice (with a suitable sequence $\theta \to 0$) is asymptotically valid even for the non-zero bias case, although in contrast to the zero bias case, a non-asymptotic result like Corollary \ref{cor2} does not hold.



\section{Additive quantile regression model}

In this section, we focus on the nonparametric additive quantile regression model:
\begin{equation}
y_{i} = g(\bm{z}_{i}) + u_{i}, \ g(\bm{z}) = \bar{c} + \sum_{k=1}^{d} g_{k}(z_{k}), \ \mathrm{P}(u_{i} \leq 0 \mid \bm{z}_{i}) = \tau,
\label{eq:model2}
\end{equation}
where $\bar{c},g_{1},\dots,g_{d}$ are unknown. Let $\mathcal{Z}_{k} \subset \mathbb{R}$ denote the support of $z_{1k}$ for each $k=1,\dots,d$. For identification, we normalize $g_{1},\dots,g_{d}$ such that
\begin{equation*}
\int_{\mathcal{Z}_{k}} g_{k}(z) dz = 0, \ k=1,\dots,d.
\end{equation*}
We allow for the possibility that $d$ is much larger than $n$. 
We assume that some of the functions are zero. Without loss of generality, we assume that $g_{d_{1}+1} \equiv 0, \dots, g_{d} \equiv 0$.
As in \cite{HHW10}, we further assume that the number of non-zero functions, $d_{1}$, is fixed.

Suppose that for each $k=1,\dots,d$ we have a set of basis functions $\{ \psi_{kj} : j=1,\dots,m \}$ on $\mathcal{Z}_{k}$ such that 
\begin{equation}
\int_{\mathcal{Z}_{k}} \psi_{kj}(z) dz = 0, \ j=1,\dots,m. \label{norm}
\end{equation}
For each $k$, we have a series approximation: $g_{k}(z) \approx \sum_{j=1}^{m} \beta_{kj}\psi_{kj}(z)$. Define $x_{iG_{1}} := 1, \bm{x}_{iG_{k}} := (\psi_{k-1,1}(z_{i,k-1}),\dots,\psi_{k-1,m}(z_{i,k-1}))'$ for $k=2,\dots,d+1$ and $\bm{x}_{i}:=(x_{iG_{1}},\bm{x}_{iG_{2}}',\dots,\bm{x}_{iG_{q}}')'$ with $q=d+1$. For $\bm{\beta} = (\beta_{0},\beta_{11},\dots,\beta_{1m},\beta_{21},\dots,\beta_{dm})' \in \mathbb{R}^{1+dm}$, we write $\beta_{G_{1}} = \beta_{0}$ and $\bm{\beta}_{G_{k}} = (\beta_{k-1,1},\dots,\beta_{k-1,m})'$ for $k=2,\dots,q=d+1$. In what follows, we follow the same notation as Section 3. 
As noted in Introduction, the group Lasso is suited to the estimation of high dimensional additive models, since selecting variable $z_{ik}$ is equivalent to make $\bm{\beta}_{G_{k}} = \bm{0}$. 

The proposed estimation method for $g$ is as follows. 
\begin{enumerate}
\item Determine a small $\theta \in (0,1)$ and a constant $c > 0$.
\item Compute $\tilde{\Lambda}(1-\theta | \bm{z}_{1}^{n})$ by simulation.
\item For $\lambda = c \tilde{\Lambda}(1-\theta | \bm{z}_{1}^{n})$, compute the group Lasso estimate:
\begin{equation*}
\hat{\bm{\beta}} := \arg \min_{\bm{\beta} \in \mathbb{R}^{1+dm}} \left[ \frac{1}{n} \sum_{i=1}^{n} \rho_{\tau}(y_{i} - \bm{x}_{i}'\bm{\beta}) + \frac{\sqrt{m} \lambda}{n} \sum_{k=2}^{q} \| \hat{\bm{\Sigma}}^{1/2}_{k} \bm{\beta}_{G_{k}} \|_{2} \right].
\end{equation*}
\item Construct $\hat{g}(\bm{z}) := \hat{\beta}_{0} + \sum_{k=1}^{d} \sum_{j=1}^{m} \hat{\beta}_{kj} \psi_{kj}(z_{k})$. 
\end{enumerate}

We wish to derive the convergence rate of $\hat{g}$. To this end, we introduce some regularity conditions. Unless otherwise stated, all parameter values are indexed by $n$. For $\nu > 0$, let $C^{\nu} (E)$ denote the set of all continuous functions $h$ on a bounded set $E \subset \mathbb{R}$ such that the $\underline{\nu}$-derivative $h^{(\underline{\nu})}$ exists and is $(\nu - \underline{\nu})$-H\"{o}lder continuous, where $\underline{\nu}$ is the greatest integer smaller than $\nu$. 

\begin{description}
\item[(D1)] $\{ (y_{i},\bm{z}_{i}')' : i=1,2,\dots \}$ are i.i.d. where the pair $(y_{1},\bm{z}_{1}')'$ satisfies the model (\ref{eq:model2}). 
\item[(D2)] The support $\mathcal{Z}$ of $\bm{z}_{1}$ is a compact subset of $\mathbb{R}^{d}$. 
\item[(D3)] Let $F(u | \bm{z})$ denote the conditional distribution function of $u_{1}$ given $\bm{z}_{1} = \bm{z}$. Assume that $F(u | \bm{z})$ has a continuously differentiable  density $f(u | \bm{z})$ such that there exist positive constants $\varrho, c_{f}, C_{f}$ and $L_{f}$ independent of $n$ such that $c_{f} \leq f(u | \bm{z}) \leq C_{f}$ on $[-\varrho, \varrho] \times \mathcal{Z}$, and $|f'(u | \bm{z})| \leq L_{f}$ on the support of $(u_{1},\bm{z}_{1}')'$.
\item[(D4)] $g_{k} \in C^{\nu}(\mathcal{Z}_{k})$ for $k=1,\dots,d_{1}$, where $\nu > 1/2$ is a fixed constant. 
\item[(D5)] $\max_{1 \leq k \leq d} \sup_{z \in \mathcal{Z}_{k}} \| (\psi_{k1}(z),\dots,\psi_{km}(z))' \|_{2} = O(m^{1/2})$. 
\item[(D6)] For each $k=1,\dots,d_{1}$, there exists a vector $(\bar{\beta}_{k1}^{0},\dots,\bar{\beta}^{0}_{km})'$ such that $\sup_{z \in \mathcal{Z}_{k}}| g_{k}(z) - \sum_{j=1}^{m} \bar{\beta}^{0}_{kj} \psi_{kj}(z) | = O(m^{-\nu})$ as $m \to \infty$.
\item[(D7)] The smallest eigenvalue of $\bm{\Sigma}_{k}$ is bounded away from zero uniformly over $(n,k)$, and the maximum eigenvalue of $\bm{\Sigma}_{k}$ is bounded uniformly over $(n,k)$.
\item[(D8)] $1 \lesssim \phi_{\min}$ and $\phi_{\max} \lesssim 1$. 
\item[(D9)] $m \asymp n^{1/(2\nu + 1)}$ and $m\log d /n \to 0$ as $n \to \infty$. 
\end{description}

Conditions (D1)-(D6) are standard in the literature. In fact, they are adapted from \cite{HL05}. We also refer to \cite{N97} for basic materials on series estimation.  
Condition (D7) is also standard. Condition (D8) is a restricted eigenvalue condition. We will give an example in which  conditions (D7) and (D8) are satisfied. Condition (D9) determines the order of $m$. It also restricts the magnitude of $d$. We allow for the possibility that $d$ is of order $o\{ \exp (\const \times n^{2 \nu/(2\nu + 1)}) \}$, which can diverge faster than $n$.

For a function $h: \mathcal{Z} \to \mathbb{R}$, define $\| h \|_{L_{2}} := \mathrm{E}[ h(\bm{z}_{1})^{2} ]^{1/2}$. 

\begin{theorem}
\label{thm2}
Assume conditions (D1)-(D9). Let  $t=t(n) \to \infty$ be a sequence  such that $t^{2} \{ n^{(1-2\nu)/(2\nu +1)} \vee ( m \log d /n) \} \to 0$. Take $\lambda = c \tilde{\Lambda}(1-\theta | \bm{z}_{1}^{n})$ where $\theta=(e \vee q^{1/m})^{-t^{2}}$ and $c$ is a positive constant. Then, we have $\| \hat{g} - g \|_{L_{2}} \lesssim_{p} t (n^{-\nu/(2\nu + 1)} \vee \sqrt{\log d/n})$.
\end{theorem}

\begin{remark}
If $\log d = O(m)$, the term $n^{-\nu/(2\nu+1)}$ is dominating, while if $m = o (\log d)$ (i.e., $d$ is faster than $\exp ( \const \times n^{\nu/(2\nu+1)} )$), the term $\sqrt{\log d/n}$ is dominating. 
\end{remark}

The proof of Theorem \ref{thm2} appears in Appendix D. The proof is basically a verification of the conditions of Corollary \ref{cor1}. 
Theorem \ref{thm2} shows that $\hat{g}$ can attain the convergence rate arbitrarily close to Stone's (1982, 1985) oracle rate $n^{-\nu/(2\nu+1)}$ in the case that $\log d = O(m)$ (which still allows for the possibility that $d$ has an exponential order in $n$). 
On the other hand, \citet[][Corollary 3.1]{HHW10} showed that for additive mean regression models, under a set of similar conditions, the group Lasso estimator has the convergence rate $n^{-\nu/(2\nu+1)} \sqrt{\log (d \vee n)}$. The latter rate is significantly slower than $n^{-\nu/(2\nu+1)}$ if $d$ has an exponential order. For instance, if $d=\exp (\const \times n^{1/(2\nu + 1)})$, $n^{-\nu/(2\nu+1)} \sqrt{\log (d \vee n)} \asymp n^{-(\nu-1/2)/(2\nu+1)}$. Although Theorem \ref{thm2} focuses on the quantile regression case, a similar result would apply to the mean regression case.

One might wonder that Corollary \ref{cor1} implies that the $\ell_{1}$-penalized quantile regression estimator (with a suitable choice of the tuning parameter) has a convergence rate like $n^{-\nu/(2\nu+1)} \sqrt{\log (d \vee n)}$. 
However, that is not the case. The problem happens when verifying condition (C8), in particular, that $\Delta \lesssim 1$. For the $\ell_{1}$-penalty case, since $p_{k} = 1$ for all $k$, to make $\Delta \lesssim 1$, 
we need to assume that $a_{\bar{\bm{\beta}}} = O(n^{-1/2})$, which requires the undersmoothing. Thus, at this moment, it is safe to say that it is not known whether the $\ell_{1}$-penalized estimator has a convergence rate like  $n^{-\nu/(2\nu+1)} \sqrt{\log (d \vee n)}$ for nonparametric additive quantile regression models. 

We give an example in which conditions (D7) and (D8) are satisfied. Suppose that $\bm{z}_{1}$ is uniformly distributed on $[0,1]^{d}$ and the basis functions are common for all $k=1,\dots,d$, i.e, $\psi_{kj} = \psi_{j}$ for all $k=1,\dots,d$ and $j=1,\dots,m$. In that case, $\mathcal{Z}_{k} = [0,1]$ for all $k=1,\dots,d$ and $\mathcal{Z} = [0,1]^{d}$. Define the $m \times m$ matrix $\bm{\Psi}$ by 
\begin{equation*}
\bm{\Psi} := \left[ \int_{0}^{1} \psi_{i}(z) \psi_{j}(z) dz \right]_{1 \leq i,j \leq m}.
\end{equation*}

\begin{lemma}
\label{lem7}
Suppose that $\bm{z}_{1}$ is uniformly distributed on $[0,1]^{d}$ and the basis functions are common for all $k=1,\dots,d$.
If the smallest eigenvalue of $\bm{\Psi}$ is bounded away from zero uniformly over $n$, and the maximum eigenvalue of $\bm{\Psi}$ is bounded uniformly over $n$, then so is $\bm{\Sigma}$.
In particular, conditions (D7) and (D8) are satisfied. 
\end{lemma}

The proof of Lemma \ref{lem7} appears in Appendix D. 
If the basis functions are orthonormal with respect to the Lebesgue measure (such as Fourier series), $\bm{\Psi}$ is the identity matrix, and the condition of Lemma \ref{lem7} is trivially satisfied.
Suitably normalized polynomial splines also satisfy the condition of Lemma \ref{lem7} \citep[see the proof of][Theorem 7]{N97}.

\section{Simulation experiments}

In this section, we report simulation results to present the practical performance of the group Lasso estimator. We compare the group Lasso estimator with the $\ell_{1}$-penalized quantile regression estimator and the unpenalized quantile regression estimator. Throughout all cases, we used the SeDuMi package in MATLAB to compute these estimates, and run $1,000$ repetitions. Note that since the SeDuMi package is based on an interior point algorithm, the number of non-zero elements in the unpenalized quantile regression estimator may be larger than the sample size. Although we have not formally discussed the model selection property of the group Lasso, for reference, we report the number of selected variables in the simulation experiments. 

We first consider the zero bias case. Let $\Phi (\cdot)$ denote the distribution function of the standard normal distribution. 
\begin{align*}
\text{Model 1}: \ &y_{i} = \bm{x}_{i}'\bar{\bm{\beta}} + e_{i}, \ x_{i1} = 1, \ \bm{x}_{i,-1} \sim N(\bm{0},\{ (0.25)^{|j-k|} \}_{j,k}), \\
&e_{i} - \Phi^{-1}(\tau)\sim N(0,1), \ \bm{x}_{i} \indep e_{i}.
\end{align*}
We took $p=501, q=101, p_{2} = \cdots = p_{100} = 5, n=200$ and $\tau \in \{ 0.25,0.5,0.75 \}$. For the group Lasso and the $\ell_{1}$-penalized quantile regression estimators, we followed the data-dependent choice of the tuning parameter discussed in Section 3.3. In each case, we took $\theta = 0.1$ and $c = 1.1$. 
For the vector $\bar{\bm{\beta}}$, we consider two cases. 
\begin{description}
\item (Case 1) $\bar{\bm{\beta}} = (1,\underbrace{1,\dots,1}_{5},0,\dots,0)'$.
\item (Case 2) $\bar{\bm{\beta}} = (1,\underbrace{\underbrace{1,0,\dots,0}_{5},\cdots,\underbrace{1,0,\dots,0}_{5}}_{\text{5 groups}},0,\dots,0)'$.
\end{description}
By the discussion following Corollary \ref{cor1}, Case 1 is favorable to the group Lasso since the ratio $p_{\bar{S}}/\| \bar{\bm{\beta}} \|_{0} = 6/6 = 1$ is small. On the other hand, Case 2 is not favorable to the group Lasso 
since the ratio $p_{\bar{S}}/\| \bar{\bm{\beta}} \|_{0} = 26/6 \approx 4.3$ is relatively large. In fact, Cases 1 and 2 are the best and worst case scenarios for the group Lasso, respectively, since in Case 1, all elements in the active group $G_{2}$ are non-zero, while in Case 2, only a single element in each active group is non-zero.

Table \ref{table1} shows the simulation results for Model 1. In Case 1, as expected, the group Lasso estimator performs the best. Its RMSE is nearly a half of that of the $\ell_{1}$-penalized quantile regression estimator, and nearly one fourth of that of the unpenalized quantile regression estimator.
However, in Case 2, the performance of the group Lasso estimator is inferior to that of the $\ell_{1}$-penalized quantile regression estimator, and little surprisingly, inferior to even that of the unpenalized quantile regression estimator for $\tau \in \{ 0.25,0.75 \}$.  In Case 2, the group Lasso estimator selects redundant models. For instance, for $\tau=0.5$, the group Lasso  selects on average about $20$ variables, while the true number of non-zero coefficients is $6$. This fact possibly worsens the performance of the group Lasso estimator. These observations are consistent with our theoretical results.

\begin{center}
Table \ref{table1} is about here.
\end{center}

We next consider nonparametric additive models. 
\begin{align*}
\text{Model 2}:  \ &y_{i} = 0.1 + \sum_{k=1}^{100} g_{k}(z_{ik})+ 0.5 \sigma (\bm{z}_{i}) e_{i}, \\
&g_{1}(z) = z, \ g_{2}(z) = \cos (\pi z), \ g_{3}(z) = e ( e^{z} - e + e^{-1}), \ g_{4} \equiv 0, \dots, g_{100} \equiv 0, \\
&\sigma^{2} (\bm{z}) = 0.7 + 0.1 z_{1}^{2} + 0.1 z_{2}^{2} + 0.1 z_{3}^{2}, \ \bm{z}_{i} \sim \text{Unif} [-1,1]^{100}, \\
&e_{i} - \Phi^{-1}(\tau) \sim N(0,1), \ \bm{z}_{i} \indep e_{i}.
\end{align*}
This model incorporates the conditional heteroscedasticity. 
We took $n \in \{ 400, 800 \}$ and $\tau \in \{ 0.25,0.5,0.75 \}$, and followed the data-dependent choice of the tuning parameter discussed in Section 3.3 with $\theta =0.2$ and $c=1$. 
We used cubic splines with $4$ equidistant knots, so $m=7$ and $p=1+dm=701$ in this case.

Table \ref{table2} shows the simulation results for Model 2. The group Lasso estimator performs significantly better than other estimators, especially when $n=800$. 
The improvement of RMSE of the group Lasso estimator over that of the $\ell_{1}$-penalized quantile regression estimator is about from $30 \%$ to $40 \%$ when $n=800$. A bit interesting point is that RMSE of the unpenalized quantile regression estimator becomes worse when $n$ increase from $400$ to $800$. This is possibly because the estimates of zero coefficients become larger when $n$ increases, and $n=800$ is not enough for estimating full $701$ coefficients. 

\begin{center}
Table \ref{table2} is about here.
\end{center}

Finally, we shall comment that in most cases, the number of selected of groups by the group Lasso is not much deviated from the truth. This fact indicates the possibility that in the quantile regression case, under suitable conditions,
the number of selected groups by the group Lasso is of the same stocahstic order as the truth. Similar results are known in the linear regression case \citep{LPTV10} and the $\ell_{1}$-penalized quantile regression case \citep{BC10}. 
\cite{BC10}'s argument depends on the specific property of the linear programming formulation of the $\ell_{1}$-penalized quantile regression problem, and is not directly applicable to the group Lasso case. The model selection property of the group Lasso in the quantile regression case is left in the future research. 

\section*{Acknowledgments}

This work was supported by the
Grant-in-Aid for Young Scientists (B) (22730179) from the JSPS.

\appendix

\section{Proof of Theorem \ref{thm1}}

\subsection{Preliminaries}

We introduce some technical results used in the proofs. The next theorem, which is due to \citet[][Theorem 4.7]{LT91}, is a Bernstein type inequality for vector valued Rademacher processes. 

\begin{theorem}[A concentration inequality for Rademacher processes]
\label{thmA1}
Let $\epsilon_{1},\dots,\epsilon_{n}$ be independent Rademacher random variables.
Let $B$ be a Banach space with norm $\| \cdot \|$ such that for some countable subset $D$ in the unit ball of $B'$ (the dual of $B$), $\| x \| = \sup_{f \in D} | f(x) |$ for all $x \in B$. 
Let $x_{1},\dots,x_{n}$ be arbitrary points in $B$. Put $Z := \| \sum_{i=1}^{n} \epsilon_{i} x_{i} \|$. Then, for every $t  > 0$, we have 
\begin{equation*}
\mathrm{P} \{  Z  > M(Z) + t \} \leq 2 \exp \{ -t^{2}/(8 \sigma^{2}) \}, \\
\end{equation*}
where $M(Z)$ is the median of $Z$ and $\sigma := \sup_{f \in D} \sqrt{ \sum_{i=1}^{n} f(x_{i})^{2}}$. If the one-sided inequality ``$Z  > M(Z) + t$'' is replaced by the two-sided inequality ``$| Z - M(Z) | > t$'', then the constant $2$ in front of the exponential term is replaced by $4$. 
\end{theorem}

An immediate corollary of Theorem \ref{thmA1} is:
\begin{corollary}
\label{corA1}
Work with the same notation as Theorem \ref{thmA1}. Then, for every $\lambda > 0$, 
\begin{equation*}
\mathrm{E}[e^{\lambda Z}] \leq 16 \exp \{ \lambda \sqrt{2 \mathrm{E}[Z^{2}]} + 4 (\lambda \sigma)^{2} \}.
\end{equation*}
\end{corollary}
\begin{proof}
Put $Z' := |Z - M(Z)|$. By Theorem \ref{thmA1}, $\mathrm{P}(Z'>t) \leq 4 \exp\{ -t^{2}/(8\sigma^{2}) \}$ for all $t > 0$.  By using the formula for the expectation of positive random variables, 
\begin{align*}
\mathrm{E}[e^{\lambda Z'}] &= \int_{0}^{\infty} \mathrm{P}(e^{\lambda Z'} > t) dt \\
&= 1 + \int_{1}^{\infty} \mathrm{P}(Z' > \log t/\lambda) dt \\
&= 1 + \lambda \int_{0}^{\infty} e^{\lambda t} \mathrm{P}(Z' > t) dt \\
&\leq 1 + 4 \lambda \int_{-\infty}^{\infty} \exp \{ \lambda t - t^{2}/(8 \sigma^{2})\} dt \\
&= 1 + 8 \sqrt{2 \pi} (\lambda \sigma) e^{2 (\lambda \sigma)^{2}}.
\end{align*}
Since $e^{x^{2}} \geq 1+x^{2} \geq x$, the right side is bounded by $(1+8 \sqrt{\pi}) e^{4 (\lambda \sigma)^{2}} \leq 16 e^{4 (\lambda \sigma)^{2}}$. 
Thus, we have $\mathrm{E}[e^{\lambda Z}] \leq e^{\lambda M(Z)} \mathrm{E}[e^{\lambda |Z-M(Z)|}] \leq 16 \exp \{ \lambda M(Z)+4 (\lambda \sigma)^{2} \}$.
The desired result now  follows from the fact that $M(Z) \leq \sqrt{2 \mathrm{E}[Z^{2}]}$.
\end{proof}

\subsection{Proof of Theorem \ref{thm1}}

In this subsection, we provide a proof of Theorem \ref{thm1}. The proof consists of series of lemmas.
We first prepare some notation.
Put $\bm{w}_{i} := (y_{i},\bm{x}_{i}')'$ and $m_{\bm{\beta}}(\bm{w}) :=  \rho_{\tau}(y - \bm{x}'\bm{\beta}) - \rho_{\tau}(y - \bm{x}'\bar{\bm{\beta}})$. 
Define $M(\bm{\beta}) := \mathrm{E}[ m_{\bm{\beta}}(\bm{w}_{1})]$. 
For a function $f : \mathbb{R}^{p+1} \to \mathbb{R}$ such that $\mathbb{E}[ | f(\bm{w}_{1}) |] < \infty$, we use the notation $\mathbb{G}_{n}f := n^{-1/2} \sum_{i=1}^{n} \{ f(\bm{w}_{i}) - \mathrm{E}[f(\bm{w}_{1})] \}$. 
Define 
\begin{align*}
&\Omega_{0} := \{ \| \hat{\bm{\Sigma}}^{1/2}_{k} - \bm{I}_{p_{k}} \| \leq 0.5, 2 \leq \forall k \leq q \}, \\
&\Lambda : = 
\max_{1 \leq k \leq q} \|  \sum_{i=1}^{n} \{ \tau - I(y_{i} \leq \bm{x}_{i}'\bar{\bm{\beta}}) \} (\hat{\bm{\Sigma}}_{k}^{-1/2}\bm{x}_{i G_{k}}/\sqrt{p_{k}}) \|_{2}, \\
&\mathcal{B}_{\delta}  :=  \{ \bm{\beta} \in \mathbb{R}^{p} : \bm{\beta}-\bar{\bm{\beta}} \in \mathbb{C}, \ \|\bm{\beta} - \bar{\bm{\beta}} \|_{2} = \delta \}, \ \text{for} \ \delta > 0, \\
&r_{*} := \frac{6 C_{f} a_{\bar{\bm{\beta}}}}{c_{f} \phi_{\min}}, 
\end{align*}
where $\hat{\bm{\Sigma}}_{k}^{-1/2}$ is interpreted as the generalized inverse of $\hat{\bm{\Sigma}}_{k}^{1/2}$ if it is singular (i.e., if $\hat{\bm{\Sigma}}_{k} = \bm{U} \bm{D} \bm{U}'$ denotes the spectral decomposition of $\hat{\bm{\Sigma}}_{k}$ where $\bm{U}$ is a $p_{k} \times p_{k}$ orthogonal matrix and $\bm{D}$ is a diagonal matrix with diagonal entries $d_{1} \geq \cdots  \geq d_{l} > 0=  d_{l+1} = \cdots = d_{p_{k}}$, then $\hat{\bm{\Sigma}}_{k}^{-1/2} = \bm{U} \diag \{ d_{1}^{-1/2},\dots, d_{l}^{-1/2}, 0, \dots, 0 \} \bm{U}'$). 
Recall that on $\Omega_{0}$, $\hat{\bm{\Sigma}}_{k}$ are nonsingular for all $k$. 

The first step of the proof is to establish that $\hat{\bm{\beta}}-\bar{\bm{\beta}} \in \mathbb{C}$ on the event $\{ \lambda \geq c_{1} \Lambda \} \cap \Omega_{0}$. We  estimate the probability of the event $\{ \lambda \geq c_{1} \Lambda \}$ in Lemma \ref{lem6}. 
\begin{lemma}
\label{lem3}
$\{ \lambda \geq c_{1} \Lambda \} \cap \Omega_{0}\subset \{ \hat{\bm{\beta}}-\bar{\bm{\beta}} \in \mathbb{C} \}$.
\end{lemma}
\begin{proof}
By convexity of the check function, we have 
\begin{equation*}
\rho_{\tau}(y_{i} - \bm{x}_{i}'\hat{\bm{\beta}}) - \rho_{\tau}(y_{i} - \bm{x}_{i}'\bar{\bm{\beta}}) \geq - \{ \tau - I(y_{i} \leq \bm{x}_{i}'\bar{\bm{\beta}}) \} \bm{x}_{i}'(\hat{\bm{\beta}}-\bar{\bm{\beta}}),
\end{equation*}
which implies that on $\Omega_{0}$, 
\begin{align*}
0 &\geq \sum_{i=1}^{n} \{ \rho_{\tau}(y_{i} - \bm{x}_{i}'\hat{\bm{\beta}}) - \rho_{\tau}(y_{i} - \bm{x}_{i}'\bar{\bm{\beta}}) \} +   \lambda \sum_{k=2}^{q}\sqrt{p_{k}}(\| \hat{\bm{\Sigma}}^{1/2}_{k}\hat{\bm{\beta}}_{G_{k}} \|_{2} - \| \hat{\bm{\Sigma}}^{1/2}_{k}\bar{\bm{\beta}}_{G_{k}}\|_{2}) \\
&\geq - \sum_{i=1}^{n} \{ \tau - I(y_{i} \leq \bm{x}_{i}'\bar{\bm{\beta}}) \} \bm{x}_{i}'(\hat{\bm{\beta}}-\bar{\bm{\beta}})   +   \lambda \sum_{k=2}^{q} \sqrt{p_{k}} (\| \hat{\bm{\Sigma}}^{1/2}_{k}\hat{\bm{\beta}}_{G_{k}} \|_{2} - \| \hat{\bm{\Sigma}}^{1/2}_{k}\bar{\bm{\beta}}_{G_{k}}\|_{2}) \\
&\geq - \sum_{k=1}^{q}  \left [ \sum_{i=1}^{n} \{ \tau - I(y_{i} \leq \bm{x}_{i}'\bar{\bm{\beta}}) \} (\hat{\bm{\Sigma}}_{k}^{-1/2}\bm{x}_{iG_{k}}/\sqrt{p_{k}}) \right ]' \sqrt{p_{k}}\hat{\bm{\Sigma}}_{k}^{1/2}(\hat{\bm{\beta}}-\bar{\bm{\beta}})_{G_{k}} \\
&\quad -  \lambda \sum_{k \in \bar{S}_{-1}}  \sqrt{p_{k}} \| \hat{\bm{\Sigma}}_{k}^{1/2} (\hat{\bm{\beta}} - \bar{\bm{\beta}})_{G_{k}} \|_{2}  + \lambda \sum_{k \in \bar{S}^{c}} \sqrt{p_{k}}  \| \hat{\bm{\Sigma}}_{k}^{1/2}\hat{\bm{\beta}}_{G_{k}} \|_{2} \\
&\geq -  (\Lambda + \lambda)  \sum_{k \in \bar{S}} \sqrt{p_{k}} \| \hat{\bm{\Sigma}}_{k}^{1/2}(\hat{\bm{\beta}}-\bar{\bm{\beta}})_{G_{k}} \|_{2} + (\lambda - \Lambda) \sum_{k \in \bar{S}^{c}} \sqrt{p_{k}}  \|\hat{\bm{\Sigma}}_{k}^{1/2} \hat{\bm{\beta}}_{G_{k}} \|_{2}.
\end{align*}
Thus, we have 
\begin{equation*}
(\lambda - \Lambda) \sum_{k \in \bar{S}^{c}} \sqrt{p_{k}}  \|\hat{\bm{\Sigma}}_{k}^{1/2} \hat{\bm{\beta}}_{G_{k}} \|_{2} \leq   (\Lambda + \lambda)  \sum_{k \in \bar{S}} \sqrt{p_{k}} \| \hat{\bm{\Sigma}}_{k}^{1/2}(\hat{\bm{\beta}}-\bar{\bm{\beta}})_{G_{k}} \|_{2}. 
\end{equation*}
On the event $\{ \lambda \geq c_{1} \Lambda \} \cap \Omega_{0}$, this inequality implies that $\hat{\bm{\beta}} - \bar{\bm{\beta}} \in \mathbb{C}$. 
\end{proof}

The next lemma relates the bound on $\| \hat{\bm{\beta}} - \bar{\bm{\beta}} \|_{2}$ to the tail behavior of $\mathbb{G}_{n}m_{\bm{\beta}}$ over $\mathbb{C}$. 
Recall the definitions of $r_{*}$ and $r^{*}$.

\begin{lemma}
\label{lem4}
For any $\delta \in [r_{*},r^{*})$, we have 
\begin{align*}
&\{ \| \hat{\bm{\beta}} - \bar{\bm{\beta}} \|_{2} \geq \delta \} \cap \{ \lambda \geq c_{1} \Lambda \} \cap \Omega_{0} \\
&\subset \left \{ \sqrt{n} \sup_{\bm{\beta} \in \mathcal{B}_{\delta}} | \mathbb{G}_{n} m_{\bm{\beta}} | > \delta \left (  \frac{1}{6} c_{f} \phi_{\min}^{2} n \delta - 1.5 \lambda \sqrt{p_{\bar{S}_{-1}}} \right ) \right \}.
\end{align*}
\end{lemma}
\begin{proof}
By convexity of the objective function, the fact that $\mathbb{C}$ is a cone and the previous lemma, on the event $\{ \| \hat{\bm{\beta}} - \bar{\bm{\beta}} \|_{2} \geq \delta \} \cap \{ \lambda \geq c_{1} \Lambda \} \cap \Omega_{0}$, there exists a vector $\bm{\beta} \in \mathcal{B}_{\delta}$ such that 
\begin{align}
0 &\geq \sum_{i=1}^{n} m_{\bm{\beta}}(\bm{z}_{i}) + \lambda \sum_{k=2}^{q}  \sqrt{p_{k}} (\| \hat{\bm{\Sigma}}_{k}^{1/2} \bm{\beta}_{G_{k}} \|_{2} - \| \hat{\bm{\Sigma}}_{k}^{1/2} \bar{\bm{\beta}}_{G_{k}}\|_{2} ) \notag \\
&\geq \sum_{i=1}^{n} m_{\bm{\beta}}(\bm{z}_{i}) + \lambda \sum_{k \in \bar{S}_{-1}}  \sqrt{p_{k}} ( \| \hat{\bm{\Sigma}}_{k}^{1/2} \bm{\beta}_{G_{k}} \|_{2} - \| \hat{\bm{\Sigma}}_{k}^{1/2} \bar{\bm{\beta}}_{G_{k}}\|_{2} ) \notag \\
&\geq \sum_{i=1}^{n} m_{\bm{\beta}}(\bm{z}_{i}) - \lambda \sum_{k \in \bar{S}_{-1}}  \sqrt{p_{k}}  \| \hat{\bm{\Sigma}}_{k}^{1/2}(\bm{\beta} - \bar{\bm{\beta}})_{G_{k}} \|_{2}. \label{eq:ineq}
\end{align}
For the first term on (\ref{eq:ineq}), we have 
\begin{align*}
\sum_{i=1}^{n} m_{\bm{\beta}}(\bm{z}_{i}) 
&=n M(\bm{\beta}) + \sqrt{n} \mathbb{G}_{n} m_{\bm{\beta}} \\
&\geq n M(\bm{\beta}) - \sup_{\bm{\beta} \in \mathcal{B}_{\delta}}| \sqrt{n} \mathbb{G}_{n} m_{\bm{\beta}} |. 
\end{align*}
Put $\bm{\alpha} := \bm{\beta} - \bar{\bm{\beta}}$. By Taylor's theorem, for $\delta \in [r_{*},r^{*})$, we have
\begin{align*}
M(\bm{\beta})
&\geq - C_{f} a_{\bar{\bm{\beta}}} \mathrm{E}[ | \bm{\alpha}'\bm{x}_{1} |] + \frac{1}{2} c_{f} \mathrm{E}[(\bm{\alpha}'\bm{x}_{1})^{2}] - \frac{1}{6} L_{f} \mathrm{E}[| \bm{\alpha}'\bm{x}_{1} |^{3} ] \\
&\geq - C_{f} a_{\bar{\bm{\beta}}} \mathrm{E}[(\bm{\alpha}'\bm{x}_{1})^{2}]^{1/2} + \frac{1}{2} c_{f} \mathrm{E}[(\bm{\alpha}'\bm{x}_{1})^{2}]  - \frac{L_{f}}{6} \frac{\mathrm{E}[| \bm{\alpha}'\bm{x}_{1} |^{3} ]}{\mathrm{E}[(\bm{\alpha}'\bm{x}_{1})^{2}]^{3/2}} \cdot \mathrm{E}[(\bm{\alpha}'\bm{x}_{1})^{2}]^{3/2} \\
&\geq \frac{1}{6} c_{f} \mathrm{E}[(\bm{\alpha}'\bm{x}_{1})^{2}] \left (1 - \frac{r_{*}}{\delta} \right )  + \frac{1}{6} c_{f} \mathrm{E}[(\bm{\alpha}'\bm{x}_{1})^{2}]  + \frac{1}{6} c_{f} \mathrm{E}[(\bm{\alpha}'\bm{x}_{1})^{2}]  \left ( 1 - \frac{ \delta}{r^{*}}  \right )  \\
&> \frac{1}{6} c_{f} \mathrm{E}[(\bm{\alpha}'\bm{x}_{1})^{2}] \\
&\geq  \frac{1}{6} c_{f} \phi_{\min}^{2} \delta^{2}, 
\end{align*}
where for the first inequality, it is useful to invoke Knight's (1998) identity: 
\begin{equation*}
\rho_{\tau}(u-v) - \rho_{\tau}(u) = - \{ \tau - I( u \leq 0 ) \}v + \int_{0}^{v} \{ I(u \leq s) - I( u \leq 0 ) \} ds. 
\end{equation*}
For the second term on (\ref{eq:ineq}), on the event $\Omega_{0}$, we have
\begin{equation*}
\lambda \sum_{k \in \bar{S}_{-1}}  \sqrt{p_{k}}  \| \hat{\bm{\Sigma}}_{k}^{1/2}(\bm{\beta} - \bar{\bm{\beta}})_{G_{k}} \|_{2}
\leq 1.5 \lambda \sum_{k \in \bar{S}_{-1}}  \sqrt{p_{k}}  \| (\bm{\beta} - \bar{\bm{\beta}})_{G_{k}} \|_{2} \leq 1.5 \delta \lambda \sqrt{p_{\bar{S}_{-1}}}.
\end{equation*}
Therefore, we obtain the desired conclusion. 
\end{proof}

We now analyze the tail probability of $\sqrt{n} | \mathbb{G}_{n} m_{\bm{\beta}} |$ over $\bm{\beta} \in \mathcal{B}_{\delta}$. 

\begin{lemma}
\label{lem5}
For any $\delta  > 0$ and $t \geq \phi_{\max} \delta \sqrt{8n}$, we have
\begin{align*}
&\mathrm{P} \left \{ \sqrt{n} \sup_{\bm{\beta} \in \mathcal{B}_{\delta}}  | \mathbb{G}_{n} m_{\bm{\beta}} | > t  \right \} \\
&\quad \leq 64 q \exp \left  [ - \left \{ t/\delta -c_{2}(1+\sqrt{p_{\bar{S}_{-1}}}) \sqrt{n} \right \}^{2} \Big/  \{ 4c_{2}\sqrt{(1+p_{\bar{S}_{-1}}/p_{\min})n} \}^{2} \right ] + 4 \gamma,
\end{align*}
where we recall that $c_{2} := 12 \sqrt{2} (c_{0} + 1)$.
\end{lemma}
\begin{proof}
Let $\epsilon_{1},\dots,\epsilon_{n}$ be independent Rademacher random variables independent of $\bm{w}_{1},\dots,\bm{w}_{n}$. Write $\mathrm{P}_{\epsilon}$ and $\mathrm{E}_{\epsilon}$ for the conditional probability and  the conditional expectation with respect to $\epsilon_{1},\dots,\epsilon_{n}$ given $\bm{w}_{1},\dots,\bm{w}_{n}$, respectively.
For $\bm{\beta} \in \mathcal{B}_{\delta}$, we have 
\begin{equation*}
\mathrm{E}[ m_{\bm{\beta}}(\bm{w}_{1})^{2} ] \leq \mathrm{E}[ \{ \bm{x}_{1}'(\bm{\beta}-\bar{\bm{\beta}}) \}^{2} ] \leq \phi_{\max}^{2} \delta^{2}.
\end{equation*}
Thus, by the symmetrization inequality \citep[][Lemma 2.3.7]{VW96}, for $t \geq \phi_{\max} \delta \sqrt{8n}$, we have 
\begin{equation*}
\mathrm{P} \left \{ \sqrt{n} \sup_{\bm{\beta} \in \mathcal{B}_{\delta}}  | \mathbb{G}_{n} m_{\bm{\beta}} | > t  \right \} \leq 4\mathrm{E} \left [ \mathrm{P}_{\epsilon} \left \{ \sup_{\bm{\beta} \in \mathcal{B}_{\delta}} \left | \sum_{i=1}^{n} \epsilon_{i} m_{\bm{\beta}}(\bm{w}_{i}) \right | > \frac{t}{4} \right\} \right].
\end{equation*}
By Markov's inequality, the probability on the right side is bounded by 
\begin{equation}
e^{-\frac{st}{4}} \mathrm{E}_{\epsilon}\left [ \exp \left \{ s \sup_{\bm{\beta} \in \mathcal{B}_{\delta}}  \left |  \sum_{i=1}^{n} \epsilon_{i} m_{\bm{\beta}}(\bm{w}_{i}) \right | \right \} \right ], \ \text{for} \ s > 0. \label{eq:laplace}
\end{equation}
Fix $\tilde{u}_{1}:=y_{1}-\bm{x}_{1}'\bar{\bm{\beta}},\dots,\tilde{u}_{n}:=y_{n}-\bm{x}_{n}'\bar{\bm{\beta}}$. Define $\varphi_{i}(\cdot) := \rho_{\tau}(\tilde{u}_{i} - \cdot) - \rho_{\tau}(\tilde{u}_{i})$ and $h_{\bm{\beta}} (\bm{x}) :=\bm{x}'\bm{\beta}$. 
Then, $m_{\bm{\beta}} (\bm{w}_{i}) = \varphi_{i}(h_{\bm{\beta}-\bar{\bm{\beta}}}(\bm{x}_{i}))$ and each $\varphi_{i}$ is a contraction, i.e., $| \varphi_{i}(a) - \varphi_{i}(b) |  \leq |a -b|$ for $a,b \in \mathbb{R}$. Thus, by the comparison theorem for Rademacher processes \citep[Theorem 4.12]{LT91}, we have 
\begin{equation*}
\mathrm{E}_{\epsilon}\left [ \exp \left \{ s \sup_{\bm{\beta} \in \mathcal{B}_{\delta}} \left |  \sum_{i=1}^{n} \epsilon_{i} m_{\bm{\beta}} (\bm{w}_{i}) \right | \right \}  \right]  
\leq \mathrm{E}_{\epsilon} \left [ \exp \left \{ 2 s  \sup_{\bm{\beta} \in \mathcal{B}_{\delta}} \left | \sum_{i=1}^{n} \epsilon_{i} h_{\bm{\beta}-\bar{\bm{\beta}}}(\bm{x}_{i}) \right | \right \}  \right ]. 
\end{equation*}
Put $Z_{k} := \| \sum_{i=1}^{n} \epsilon_{i}\bm{x}_{iG_{k}}/\sqrt{p_{k}} \|_{2}$. For $\bm{\beta} \in \mathcal{B}_{\delta}$, we have 
\begin{align*}
\left |  \sum_{i=1}^{n} \epsilon_{i} h_{\bm{\beta}-\bar{\bm{\beta}}}(\bm{x}_{i}) \right| &= \left |  \sum_{k=1}^{q}  \sqrt{p_{k}} (\bm{\beta} - \bar{\bm{\beta}})_{G_{k}}' \left \{ \sum_{i=1}^{n} \epsilon_{i} \bm{x}_{iG_{k}}/\sqrt{p_{k}} \right \} \right | \\
&\leq  \sum_{k \in \bar{S}} \sqrt{p_{k}} \| (\bm{\beta} - \bar{\bm{\beta}})_{G_{k}} \|_{2} Z_{k} + ( \max_{k \in \bar{S}^{c}} Z_{k}  ) \sum_{k \in \bar{S}^{c}} \sqrt{p_{k}}  \| (\bm{\beta} - \bar{\bm{\beta}})_{G_{k}} \|_{2}  \\
&\leq \sum_{k \in \bar{S}} \sqrt{p_{k}} \| (\bm{\beta} - \bar{\bm{\beta}})_{G_{k}} \|_{2} Z_{k} + ( \max_{k \in \bar{S}^{c}} Z_{k}  )c_{0} \sum_{k \in \bar{S}} \sqrt{p_{k}} \| (\bm{\beta} - \bar{\bm{\beta}})_{G_{k}} \|_{2} \\
&\leq \delta (c_{0} + 1) ( \max_{1 \leq k \leq q} Z_{k} + \sqrt{p_{\bar{S}_{-1}}} \max_{2 \leq k \leq q} Z_{k} ). 
\end{align*}
Thus, by the Cauchy-Schwarz inequality, 
\begin{align*}
\mathrm{E}_{\epsilon} &\left [ \exp \left \{ 2 s   \sup_{\bm{\beta} \in \mathcal{B}_{\delta}} \left | \sum_{i=1}^{n} \epsilon_{i} h_{\bm{\beta}-\bar{\bm{\beta}}}(\bm{x}_{i}) \right |  \right \} \right ] 
\leq \mathrm{E}_{\epsilon} \left [ \exp \left \{ 2  s \delta (c_{0} + 1) ( \max_{1 \leq k \leq q} Z_{k} + \sqrt{p_{\bar{S}_{-1}}} \max_{2 \leq k \leq q} Z_{k} )  \right \} \right ]  \\
&\leq \left \{  \mathrm{E}_{\epsilon} \left [ \exp \left \{ 4 s \delta (c_{0} + 1) \max_{1 \leq k \leq q} Z_{k} \right \} \right ] \right \}^{1/2} 
\cdot \left \{  \mathrm{E}_{\epsilon} \left [ \exp \left \{ 4 s \delta (c_{0} + 1) \sqrt{p_{\bar{S}_{-1}}}  \max_{2 \leq k \leq q} Z_{k}  \right \} \right ]  \right \}^{1/2}  \\
&\leq \left \{ \sum_{k=1}^{q} \mathrm{E}_{\epsilon} \left [ \exp \left \{ 4 s \delta (c_{0} + 1) Z_{k}  \right \} \right ] \right \}^{1/2}
\cdot \left \{ \sum_{k=2}^{q} \mathrm{E}_{\epsilon} \left [ \exp \left \{ 4 s \delta (c_{0} + 1) \sqrt{p_{\bar{S}_{-1}}} Z_{k}  \right \} \right ] \right \}^{1/2}.
\end{align*}
Put  $a :=  4 s \delta (c_{0} + 1)$. By Corollary \ref{corA1}, on $\Omega_{0}$, we have 
\begin{equation*}
\mathrm{E}_{\epsilon} [ \exp ( a Z_{k} ) ] \leq 16 \exp \{ 1.5 a \sqrt{2 n}   + 9a^{2}n/p_{k} \},
\end{equation*}
where we have used the fact that on $\Omega_{0}$, 
\begin{align*}
&\mathrm{E}[ Z_{k}^{2} ] = p_{k}^{-1} \sum_{i=1}^{n} \| \bm{x}_{iG_{k}} \|_{2}^{2} = p_{k}^{-1} n \tr \hat{\bm{\Sigma}}_{k} \leq (1.5)^{2} np_{k}^{-1} \tr \bm{I}_{p_{k}} = (1.5)^{2} n, \\
\intertext{and} 
&\sup_{\bm{\alpha} \in \mathbb{S}^{p_{k}-1}} \sum_{i=1}^{n} (\bm{\alpha}'\bm{x}_{iG_{k}}/\sqrt{p_{k}})^{2} = p_{k}^{-1} n \| \hat{\bm{\Sigma}}_{k} \| \leq  (1.5)^{2} p_{k}^{-1} n.
\end{align*} 
Similarly, on $\Omega_{0}$, we have 
$\mathrm{E}_{\epsilon} [ \exp \{  a \sqrt{p_{\bar{S}_{-1}}}Z_{k} \} ] \leq 16 \exp \{ 1.5 a  \sqrt{2 p_{\bar{S}_{-1}} n}  + 9a^{2}n p_{\bar{S}_{-1}}/p_{k} \}$. Therefore, on $\Omega_{0}$, the right side on (\ref{eq:laplace}) is bounded by 
\begin{align*}
&16 q \exp \{ 0.75a (1+\sqrt{p_{\bar{S}_{-1}}} ) \sqrt{2n} + 4.5 a^{2}n(1+p_{\bar{S}_{-1}}/p_{\min} ) - st/4\} \\
&= 16 q \exp \{ 3s \delta (c_{0} + 1) (1+\sqrt{p_{\bar{S}_{-1}}}) \sqrt{2n} + s^{2} \{ 6 \sqrt{2} \delta (c_{0} + 1) \sqrt{n(1+p_{\bar{S}_{-1}}/p_{\min} )} \}^{2} -st/4\} \\
&= 16 q \exp \{ -(t-b)s/4 + cs^{2} \},
\end{align*}
where $b:= 12\delta (c_{0} + 1) (1+\sqrt{p_{\bar{S}_{-1}}} ) \sqrt{2n}$ and $c := \{ 6 \sqrt{2} \delta (c_{0} + 1) \sqrt{n(1+p_{\bar{S}_{-1}}/p_{\min})} \}^{2}$. Minimizing the right side with respect to $s > 0$, on $\Omega_{0}$, we have
\begin{equation*}
\mathrm{P}_{\epsilon} \left \{ \sup_{\bm{\beta} \in \mathcal{B}_{\delta}} \left | \sum_{i=1}^{n} \epsilon_{i} m_{\bm{\beta}}(\bm{w}_{i}) \right | > \frac{t}{4} \right\} \leq 16 q \exp \{ -(t-b)^{2}/(64c) \}. 
\end{equation*}
Therefore, we obtain the desired conclusion. 
\end{proof}

It remains to estimate $\mathrm{P}(\Lambda > c_{1}^{-1} \lambda )$. Recall that $\bm{z}_{1}^{n}:= \{ \bm{z}_{1},\dots,\bm{z}_{n} \}$.

\begin{lemma}
\label{lem6}
For any $t_{1} > 0$ and $t_{2} > 0$, 
\begin{align*}
&\mathrm{P}\{ \Lambda > (4\sqrt{2n}+\sqrt{n}\Delta + t_{1} + t_{2}) \mid \bm{z}_{1}^{n} \} \\
&\leq 2 \exp \{ -t_{1}^{2}/(2n) \} + 16 (q-1) \exp \{ -p_{\min}t_{2}^{2}/(128n)\}.
\end{align*}
\end{lemma}
\begin{proof}
Put $\check{\bm{x}}_{ik} := \hat{\bm{\Sigma}}_{k}^{-1/2}\bm{x}_{i G_{k}}/\sqrt{p_{k}}$. 
Observe that
\begin{align*}
\Lambda &\leq \max_{1 \leq k \leq q} \|  \sum_{i=1}^{n} \{ \tau - \mathrm{P}(y_{i} \leq \bm{x}_{i}'\bar{\bm{\beta}} | \bm{x}_{iG_{k}}) \}\check{\bm{x}}_{ik} \|_{2} \\
&\quad + \max_{1 \leq k \leq q} \|  \sum_{i=1}^{n} \{ \mathrm{P}(y_{i} \leq \bm{x}_{i}'\bar{\bm{\beta}} | \bm{x}_{iG_{k}}) - I(y_{i} \leq \bm{x}_{i}'\bar{\bm{\beta}}) \}\check{\bm{x}}_{ik}  \|_{2} \\
&=:\Lambda_{1} + \Lambda_{2}. 
\end{align*}

We first analyze $\Lambda_{1}$. For $2 \leq k \leq q$, we have
\begin{align*}
\|  \sum_{i=1}^{n} \{ \tau - \mathrm{P}(y_{i} \leq \bm{x}_{i}'\bar{\bm{\beta}} | \bm{x}_{iG_{k}}) \}\check{\bm{x}}_{ik} \|^{2}_{2} &= \sup_{\bm{\alpha} \in \mathbb{S}^{p_{k}-1}} | \sum_{i=1}^{n} \{ \tau - \mathrm{P}(y_{i} \leq \bm{x}_{i}'\bar{\bm{\beta}} | \bm{x}_{iG_{k}}) \} \bm{\alpha}'\check{\bm{x}}_{ik} |^{2} \\
&\leq  \sum_{i=1}^{n} \{ \tau - \mathrm{P}(y_{i} \leq \bm{x}_{i}'\bar{\bm{\beta}} | \bm{x}_{iG_{k}}) \}^{2} \sup_{\bm{\alpha} \in \mathbb{S}^{p_{k}-1}} \sum_{i=1}^{n} (\bm{\alpha}'\check{\bm{x}}_{ik} )^{2} \\
&\leq n^{2} C^{2}_{f} a^{2}_{\bar{\bm{\beta}}} /p_{k},
\end{align*}
where the second inequality is due to the Cauchy-Schwarz inequality. 
For $k=1$, by condition (C4) and Taylor's theorem, 
\begin{equation*}
|  \sum_{i=1}^{n} \{ \tau - \mathrm{P}(y_{i} \leq \bm{x}_{i}'\bar{\bm{\beta}} | x_{iG_{1}}) \}\check{x}_{iG_{1}} | \leq n | \tau - \mathrm{P}(y_{1} \leq \bm{x}_{1}'\bar{\bm{\beta}}) | \leq \frac{nL_{f}}{2} a_{\bar{\bm{\beta}}}^{2}, 
\end{equation*}
which implies that $\Lambda_{1} \leq n L_{f} a_{\bar{\bm{\beta}}}^{2}/2 \vee nC_{f} a_{\bar{\bm{\beta}}} /\sqrt{p_{\min}} = \sqrt{n} \Delta$.  

It remains to estimate $\Lambda_{2}$. Put $\eta_{ik} := \mathrm{P}(y_{i} \leq \bm{x}_{i}'\bar{\bm{\beta}} | \bm{x}_{iG_{k}}) - I(y_{i} \leq \bm{x}_{i}'\bar{\bm{\beta}})$. Observe that for $t_{1} > 0$ and $t_{2} > 0$, 
\begin{equation*}
\mathrm{P}(\Lambda_{2} > t_{1} + t_{2} \mid \bm{z}_{1}^{n}) \leq \mathrm{P}\left \{ \left |  \sum_{i=1}^{n} \eta_{i1} \right | > t_{1} \ \Bigg | \ \bm{z}_{1}^{n} \right \} 
 + (q-1) \max_{2 \leq k \leq q} \mathrm{P}\left \{ \left \|  \sum_{i=1}^{n} \eta_{ik} \check{\bm{x}}_{ik}   \right \|_{2} > t_{2} \ \Bigg | \ \bm{z}_{1}^{n} \right \}.
\end{equation*}
Since $| \eta_{i1} | \leq 1$, by Hoeffding's inequality, the first term on the right side is bounded by $2 \exp \{ -t_{1}^{2}/(2n) \}$. 
It remains to estimate the second term. Fix $2 \leq k \leq q$. Recall that for $\bm{x} \in \mathbb{R}^{p_{k}}$, $\| \bm{x} \|_{2} = \sup_{\bm{\alpha} \in \mathbb{S}^{p_{k}-1}} \bm{\alpha}'\bm{x}$. Let $\epsilon_{1},\dots,\epsilon_{n}$ be independent Rademacher random variables independent of $\bm{z}^{n}_{1}$. By using the symmetrization inequality \citep[][Lemma 2.3.7]{VW96}, for $t_{2} \geq \sqrt{8n/p_{k}}$, 
\begin{equation*}
\mathrm{P}\left \{ \left \|  \sum_{i=1}^{n} \eta_{ik} \check{\bm{x}}_{ik}  \right\|_{2} > t_{2} \ \Bigg | \ \bm{z}_{1}^{n} \right \} 
\leq 4 \mathrm{E}\left [ \mathrm{P}_{\epsilon} \left \{ \left \|  \sum_{i=1}^{n} \epsilon_{i} \eta_{ik} \check{\bm{x}}_{ik}  \right \|_{2} > \frac{t_{2}}{4}  \right \} \ \Bigg | \ \bm{z}_{1}^{n}  \right ]. 
\end{equation*}
Since $| \eta_{ik} | \leq 1$, by the contraction theorem for Rademacher processes \citep[][Theorem 4.4]{LT91}, 
the probability on the right side is bounded by 
\begin{equation*}
2\mathrm{P}_{\epsilon} \left \{ \left \|  \sum_{i=1}^{n} \epsilon_{i} \check{\bm{x}}_{ik}   \right \|_{2} > \frac{t_{2}}{4} \right \}.
\end{equation*}
The desired result now follows from the concentration inequality for Rademacher processes (see Theorem \ref{thmA1}).
\end{proof}

\begin{proof}[Proof of Theorem \ref{thm1}]
Take $\lambda = c_{1} \lambda_{A}$, 
\begin{align*}
\delta &= \left [ \frac{6}{c_{f} \phi_{\min}^{2}} \left ( \epsilon^{*}_{B} \vee \sqrt{\frac{8\phi_{\max}^{2} }{n}} + \frac{1.5 \lambda \sqrt{p_{\bar{S}_{-1}}}}{n}   \right ) \right ] \vee \frac{6 C_{f}a_{\bar{\bm{\beta}}}}{c_{f} \phi_{\min}}, \\
t &= \delta \left ( \frac{1}{6} c_{f} \phi_{\min}^{2} n \delta - 1.5 \lambda  \sqrt{p_{\bar{S}_{-1}}} \right ).
\end{align*}
Since $\delta \in [r_{*},r^{*})$, by Lemma \ref{lem4}, we have
\begin{equation*}
\mathrm{P}[ \{ \| \hat{\bm{\beta}} - \bar{\bm{\beta}} \|_{2} \geq \delta \} \cap \{ \lambda \geq c_{1} \Lambda \} \cap \Omega_{0} ] \leq
\mathrm{P} \{ \sqrt{n} \sup_{\bm{\beta} \in \mathcal{B}_{\delta}} | \mathbb{G}_{n} m_{\bm{\beta}} | > t \}.
\end{equation*}
The present choice of $t$ ensures that $t \geq \phi_{\max} \delta \sqrt{8n}$ and
\begin{equation*}
t/\delta - c_{2}(1+\sqrt{p_{\bar{S}_{-1}}})\sqrt{n}  \geq B \sqrt{\log q} \cdot 4 c_{2} \sqrt{(1+ p_{\bar{S}_{-1}}/p_{\min})n},
\end{equation*}
which implies by Lemma \ref{lem5} that
\begin{equation*}
\mathrm{P} \{ \sqrt{n} \sup_{\bm{\beta} \in \mathcal{B}_{\delta}} | \mathbb{G}_{n} m_{\bm{\beta}} | > t \} \leq 64 q^{1-B^{2}} + 4 \gamma.
\end{equation*}
Thus, we have 
\begin{equation*}
\mathrm{P} \{ \| \hat{\bm{\beta}} - \bar{\bm{\beta}} \|_{2} \geq \delta \} \leq \mathrm{P} ( \Lambda > c_{1}^{-1} \lambda ) + 64 q^{1-B^{2}} + 5 \gamma, 
\end{equation*}
where by Lemma \ref{lem6}, $\mathrm{P} ( \Lambda > c_{1}^{-1} \lambda ) = \mathrm{P} ( \Lambda > \lambda_{A} ) \leq 2 e^{-A_{1}^{2}/2} + 16 q^{1-A_{2}^{2}/128}$. Therefore, we obtain the desired conclusion. 
\end{proof}

\section{Proofs of Lemmas \ref{lem1} and \ref{lem2}}

\begin{proof}[Proof of Lemma \ref{lem1}]
It suffices to show that $\max_{2 \leq k \leq q} \| \hat{\bm{\Sigma}}_{k} - \bm{I}_{p_{k}} \| \stackrel{p}{\to} 0$. 
Fix $2 \leq k \leq q$ for a while.
Define $h_{\bm{\alpha}}(\bm{x}):=(\bm{\alpha}'\bm{x}_{G_{k}})^{2}$ for $\bm{x} \in \mathbb{R}^{p}$ and $\bm{\alpha} \in \mathbb{S}^{p_{k}-1}$.
Observe that 
\begin{equation*}
\| \hat{\bm{\Sigma}}_{k} - \bm{I}_{p_{k}} \| = \frac{1}{n} \sup_{\bm{\alpha} \in \mathbb{S}^{p_{k}-1}} \left | \sum_{i=1}^{n} \{ h_{\bm{\alpha}}(\bm{x}_{i}) - \mathrm{E}[h_{\bm{\alpha}}(\bm{x}_{1})] \} \right | =: n^{-1} Z.
\end{equation*}
By Bousquet's (2002) version of Talagrand's (1996) inequality, for all $t > 0$, 
\begin{equation*}
\mathrm{P} \left \{  Z \geq \mathrm{E}[Z] + t K \sqrt{2(n p_{k} + 4p_{k} \mathrm{E}[Z])} + \frac{2t^{2} K^{2} p_{k}}{3} \right \} \leq  e^{-t^{2}}, 
\end{equation*}
where we have used the fact that $| h_{\bm{\alpha}}(\bm{x}_{1}) | \leq K^{2}p_{k}$ and $\Var (h_{\bm{\alpha}}(\bm{x}_{1})) \leq \mathrm{E}[(\bm{\alpha}'\bm{x}_{1G_{k}})^{4}] \leq K^{2} p_{k} \mathrm{E}[(\bm{\alpha}'\bm{x}_{1G_{k}})^{2}] = K^{2}p_{k}$. We wish to estimate $\mathrm{E}[Z]$. 
By Theorem 1 of \cite{R99}, there exists a universal constant $L$ such that
\begin{equation*}
\mathrm{E}[Z] \leq  L \sqrt{n \log (p_{k} \vee e)} (\mathrm{E}[ \| \bm{x}_{1G_{k}} \|_{2}^{\log n} ])^{1/\log n}, 
\end{equation*}
provided that the last expression is smaller than $n$ (an explicit value of the constant $L$ and an elementary proof of Rudelson's inequality are given in \cite{O10}). Since $\| \bm{x}_{1G_{k}} \|_{2} \leq K \sqrt{p_{k}}$, the last expression is bounded by $KL \sqrt{n p_{\max}\log (p_{\max} \vee e)} =: U_{n}$. The present hypothesis ensures that $U_{n}$ is of order $o(n)$ and hence there exists a positive integer $n_{1}$ (independent of $k$) such that 
$U_{n}$ is smaller than $n$ for all $n \geq n_{1}$. Let $n \geq n_{1}$. Then, for any $t > 0$, with probability at least $1-e^{-t^{2}}$, 
\begin{equation*}
Z \leq U_{n} + tK\sqrt{2p_{\max} (n  + 4 U_{n})} + \frac{2t^{2}K^{2}p_{\max}}{3} =: \bar{U}_{n}(t).
\end{equation*}
Take $t=\sqrt{2 \log (q \vee n)}$. Then, with probability at least $1-q^{-2} \wedge n^{-2}$, we have $Z \leq \bar{U}_{n}(t)$. 
Since, by the present hypothesis that $p_{\max}\log (q \vee n)/n \to 0$, $\bar{U}_{n}(t)$ is of order $o(n)$ and independent of $k$, this implies  that, by the union bound, $\max_{2 \leq k \leq q} \| \hat{\bm{\Sigma}}_{k} - \bm{I}_{p_{k}} \| \stackrel{p}{\to} 0$. 
\end{proof}

\begin{proof}[Proof of Lemma \ref{lem2}]
The proof is based on Corollary 2.7 of \cite{MPT-J07}. Define the set 
\begin{equation*}
\mathcal{A} := \bigcup_{k=2}^{q} \{ \bm{\Sigma}^{1/2} \bm{\alpha} / \|  \bm{\Sigma}^{1/2} \bm{\alpha} \|_{2}  : \bm{\alpha} \in \mathbb{S}^{p-1}, \bm{\alpha}_{G_{l}} = \bm{0}, \ \forall l \neq k \} \subset \mathbb{S}^{p-1},
\end{equation*}
Let $L(\cdot)$ denote an isonormal Gaussian process on $\mathbb{R}^{p}$ with the respect to the Euclidean norm \citep[see][Chapter 1 for basic materials on isonormal Gaussian processes]{D99}. Write $| L(\mathcal{A}) | := \sup_{\bm{\alpha} \in \mathcal{A}} | L(\bm{\alpha}) |$. Pick any $\epsilon > 0$. In what follows, $c$ and $C$ denote some constants independent of $n$ and $\epsilon$. Their values may change from  line to line. Observe that $ \|  \bm{\Sigma}^{1/2} \bm{\alpha} \|_{2} = \| \bm{\Sigma}_{k}^{1/2} \bm{\alpha}_{G_{k}} \|_{2} = \| \bm{\alpha}_{G_{k}} \|$=1 for $\bm{\alpha} \in \mathbb{S}^{p-1}$ such that $\bm{\alpha}_{G_{l}} = \bm{0}$ for all $l \neq k$. By Corollary 2.7 of \cite{MPT-J07}, as long as $n \geq C \mathrm{E}[| L(\mathcal{A}) |]^{2}/\epsilon^{2}$, with probability at least $1-\exp ( - c\epsilon^{2} n )$, 
\begin{equation*}
1-\epsilon \leq \frac{1}{n} \sum_{i=1}^{n} (\bm{\alpha}'\tilde{\bm{x}}_{i})^{2} \leq 1-\epsilon, \ \forall \bm{\alpha} \in \mathcal{A},
\end{equation*}
which implies that 
\begin{equation}
1-\epsilon \leq \|  \hat{\bm{\Sigma}}^{1/2}_{k} \bm{\alpha} \|_{2}^{2} \leq 1-\epsilon, \ \forall \bm{\alpha} \in \mathbb{S}^{p_{k}-1}, 2 \leq \forall k \leq q. \label{eq:ineq2}
\end{equation}

We wish to estimate $\mathrm{E}[| L(\mathcal{A}) |]$. Define $\mathcal{U}_{k} := \{ \bm{\alpha} \in \mathbb{R}^{p} : \| \bm{\alpha} \|_{2} \leq 1, \bm{\alpha}_{G_{l}} = \bm{0}, \forall l \neq k \}$ for $k=2,\dots,q$. Recall that for $\bm{\alpha} \in \mathcal{U}_{k} \cap \mathbb{S}^{p-1}$, $\| \bm{\Sigma}^{1/2} \bm{\alpha} \|_{2} = \| \bm{\alpha}_{G_{k}} \|_{2} = 1$, so that $\mathcal{A} \subset \{ \bm{\Sigma}^{1/2} \bm{\alpha} : \bm{\alpha} \in \bigcup_{k=2}^{q} \mathcal{U}_{k} \}$. 
On $\mathbb{R}^{p}$, a version of $L(\cdot)$ is given by $L(\bm{\alpha}) = \bm{\alpha}'\bm{v}$ where $\bm{v} \sim N(\bm{0},\bm{I}_{p})$. Thus, we have
\begin{align*}
\mathrm{E}[ | L(\mathcal{A}) |] &\leq \mathrm{E}[ \sup_{\bm{\alpha} \in \cup_{k=2}^{q} \mathcal{U}_{k}} | L(\bm{\Sigma}^{1/2} \bm{\alpha}) | ] \\
&= \mathrm{E}[ \sup_{\bm{\alpha} \in \cup_{k=2}^{q} \mathcal{U}_{k}}| \bm{\alpha}'(\bm{\Sigma}^{1/2} \bm{v}) |].
\end{align*}
For each $k$, it is shown by a standard argument that there exists a $1/2$-cover $\Pi_{k}$ of $\mathcal{U}_{k}$ such that $\Pi_{k} \subset \mathcal{U}_{k}, \ \mathcal{U}_{k} \subset 2 \conv \Pi_{k}$ and $| \Pi_{k} | \leq 5^{p_{k}}$ \citep[$\conv \Pi_{k}$ stands for the convex hull of $\Pi_{k}$, and $2 \conv \Pi_{k} = \{ 2 \bm{\alpha} : \bm{\alpha} \in \conv \Pi_{k} \}$; for this result, cf.][]{MPT-J08}. 
Then, $\Pi := \bigcup_{k=2}^{q} \Pi_{k}$ is a $1/2$-cover of $\bigcup_{k=2}^{q} \mathcal{U}_{k}$ such that $\Pi \subset \bigcup_{k=2}^{q} \mathcal{U}_{k}, \ \bigcup_{k=2}^{q} \mathcal{U}_{k} \subset 2 \conv \Pi$ and 
$| \Pi | \leq (q-1) 5^{p_{\max}}$. By a maximal inequality for Gaussian random variables \citep[cf.][Eq. (3.13)]{LT91}, we have
\begin{align*}
\mathrm{E}[ \sup_{\bm{\alpha} \in \cup_{k=2}^{q} \mathcal{U}_{k}}| \bm{\alpha}'(\bm{\Sigma}^{1/2} \bm{v}) |] &\leq \mathrm{E}[ \sup_{\bm{\alpha} \in 2\conv \Pi}| \bm{\alpha}'(\bm{\Sigma}^{1/2} \bm{v}) |] \\
&=2 \mathrm{E}[ \max_{\bm{\alpha} \in \Pi} | \bm{\alpha}'(\bm{\Sigma}^{1/2} \bm{v}) |] \\
&\leq C \sqrt{p_{\max} \vee \log q} \max_{\bm{\alpha} \in \Pi} \| \bm{\Sigma}^{1/2} \bm{\alpha} \|_{2} \\
&\leq C \sqrt{p_{\max} \vee \log q}.
\end{align*}
Therefore, by the present hypothesis that $(p_{\max} \vee \log q)/n \to 0$, the inequality (\ref{eq:ineq2}) holds with probability approaching one. This implies the desired conclusion. 
\end{proof}

\section{Proof of Theorem \ref{thm3}}

The second assertion follows from the first assertion  and a careful examination of the proof of Theorem \ref{thm1}. 
Thus, we concentrate on showing the first assertion. 
Take $A_{1} = \sqrt{2\{t^{2}\log (e \vee q^{1/p_{\min}})+\log 4\}}$ and 
$A_{2} = \sqrt{128[ \{t^{2}\log (e \vee q^{1/p_{\min}}) + \log 32\}/\log q + 1]}$ so that $2e^{-A_{1}^{2}/2} = (e \vee q^{1/p_{\min}})^{-t^{2}}/2$ and $16q^{1-A_{2}^{2}/128} =  (e \vee q^{1/p_{\min}})^{-t^{2}}/2$. Since $2e^{-A_{1}^{2}/2} + 16q^{1-A_{2}^{2}/128} =  (e \vee q^{1/p_{\min}})^{-t^{2}}/2+  (e \vee q^{1/p_{\min}})^{-t^{2}}/2 = (e \vee q^{1/p_{\min}})^{-t^{2}} = \theta$, by Lemma \ref{lem6}, we have 
\begin{equation*}
\tilde{\Lambda}(1-\theta | \bm{z}_{1}^{n}) \leq (4\sqrt{2}+A_{1})\sqrt{n} + A_{2} \sqrt{n\log q/p_{\min}} \lesssim t \sqrt{n}(1+\sqrt{\log q/p_{\min}}).
\end{equation*}

We wish to establish a lower bound on $\tilde{\Lambda}(1-\theta | \bm{z}_{1}^{n})$.
By definition, 
\begin{equation*}
\tilde{\Lambda} \geq \left | \sum_{i=1}^{n} \{ \tau - I(u_{i} \leq 0) \} \right|.
\end{equation*}
We use the minorization inequality of \citet[][Theorem 5.2.2]{S74}: let $\{ X_{i}, i \geq 1 \}$ denote a sequence of independent random variables with zero mean and finite variance, and let $S_{n} = \sum_{i=1}^{n} X_{i}$ and $s_{n}^{2} = \sum_{i=1}^{n} \mathrm{E}[X_{i}^{2}]$. Suppose that there exists a positive constant $c$ such that $| X_{i} | \leq c s_{n}$ for all $1 \leq i  \leq n$. Then, for any $\zeta > 0$,  there exist positive constants $\epsilon (\zeta)$ and $\pi (\zeta)$ such that if $\epsilon \geq \epsilon (\zeta)$ and $\epsilon c \leq \pi (\zeta)$, then 
\begin{equation*}
\mathrm{P}(S_{n}/s_{n} > \epsilon) \geq \exp \{ -(\epsilon^{2}/2)(1+\zeta) \}.
\end{equation*}
Take $X_{i} = \tau - I(u_{i} \leq 0)$ and $\zeta = 1$. Then, $s_{n}^{2} = n \tau (1-\tau)$, and since $| X_{i} | \leq 1$, $c=1/s_{n} \lesssim n^{-1/2}$. Let $\epsilon = t \sqrt{\log (e \vee q^{1/p_{\min}})}$. 
Since $\epsilon$ diverges, $\epsilon \geq \epsilon (1)$ for large $n$, and $\epsilon c \lesssim n^{-1/2} t \sqrt{\log (e \vee q^{1/p_{\min}})} = o(1)$, which ensures that $\epsilon c \leq \pi (1)$ for large $n$.
Therefore, for large $n$, 
\begin{align*}
\mathrm{P} \left \{ \tilde{\Lambda}  > t \sqrt{\tau (1-\tau)n\log (e \vee q^{1/p_{\min}})} \ \Big | \ \bm{z}_{1}^{n} \right \} &\geq \mathrm{P} ( S_{n} /s_{n} > \epsilon ) \\
&\geq \exp ( -\epsilon^{2} ) \\
&=(e \vee q^{1/p_{\min}})^{-t^{2}} = \theta.
\end{align*}
Thus, we have 
\begin{equation*}
\tilde{\Lambda}(1-\theta | \bm{z}_{1}^{n}) \geq t \sqrt{\tau (1-\tau)n\log (e \vee q^{1/p_{\min}})}.
\end{equation*}
Therefore, we obtain the desired conclusion. \qed

\section{Proofs of Theorem \ref{thm2} and Lemma \ref{lem7}}

Recall that $g_{\bm{\beta}}(\bm{z}) := \beta_{0} + \sum_{k=1}^{d} \sum_{j=1}^{m} \beta_{kj} \psi_{kj}(z_{k})$ for $\bm{\beta} = (\beta_{0},\beta_{11},\dots,\beta_{1m},\beta_{21},\dots,\beta_{dm})'$. 

\begin{proof}[Proof of Theorem \ref{thm2}]

In view of the discussion on condition (C3), under condition (D7), it does not lose any generality to assume that $\bm{\Sigma}_{k} = \bm{I}_{m}$ for all $k \geq 2$. For $k=d_{1}+1,\dots,d$, let $(\bar{\beta}^{0}_{k1},\dots,\bar{\beta}^{0}_{km})' := \bm{0}_{m}$. Recall condition (D6). Define $\bar{\bm{\beta}}^{0} := (\bar{\beta}^{0}_{0},\bar{\beta}^{0}_{11},\dots,\bar{\beta}^{0}_{1m},\bar{\beta}^{0}_{21},\dots,\bar{\beta}^{0}_{dm})'$. By condition (D6) and the fact that $d_{1}$ is fixed, we have $a_{\bar{\bm{\beta}}^{0}} := \sup_{\bm{z}\in\mathcal{Z}} | g(\bm{z}) - g_{\bar{\bm{\beta}}^{0}}(\bm{z}) | = O(m^{-\nu})$.  By Lemma 3.1, there exists a vector $\bar{\bm{\beta}} \in \mathbb{R}^{p}$ such that $\mathrm{E}[ f(0 | \bm{z}_{1}) \{ g(\bm{z}_{1}) - g_{\bar{\bm{\beta}}}(\bm{z}_{1}) \}] = 0$, $\bar{\bm{\beta}}_{G_{k}} = \bm{0}$ for all $k=d_{1}+2,\dots,q=d+1$ and $a_{\bar{\bm{\beta}}} \lesssim a_{\bar{\bm{\beta}}^{0}}$. 

By Theorem \ref{thm3}, it suffices to check the conditions of Corollary \ref{cor1} (i) with $p_{2} = \cdots = p_{k} =m, q=1+d, p=1+dm$ and $| \bar{S} | = 1+d_{1}$. It is not difficult to see that conditions (C1)-(C6) are satisfied. To see that condition (C7) is satisfied, recall Lemma \ref{lem1}. By condition (D5), condition (C7) is satisfied if $m \log (d \vee n)/n \to 0$, but this is ensured by condition (D9). 
To see that condition (C8) is satisfied, recall that $a_{\bar{\bm{\beta}}} = O(m^{-\nu})$. On the other hand, 
\begin{equation*}
\frac{1}{\sqrt{n}} \wedge \frac{p_{\min}}{n} \wedge \frac{p_{\bar{S}}}{n} \left ( 1+\frac{\log q}{p_{\min}} \right ) = \frac{1}{\sqrt{n}} \wedge \frac{m}{n} \wedge \frac{1+d_{1} m}{n} \left (1+\frac{\log (1+d)}{m} \right ) \gtrsim n^{-2\nu/(2\nu+1)}.
\end{equation*}
Thus, condition (C8) is also satisfied. Therefore, we have 
\begin{equation}
\| \hat{\bm{\beta}} - \bar{\bm{\beta}} \|_{2} \lesssim_{p} t \sqrt{\frac{1+d_{1}m}{n} \left ( 1+\frac{\log(1+d)}{m} \right )} \asymp t (n^{-\nu/(2\nu+1)} \vee \sqrt{\log d/n}),
\label{conv}
\end{equation}
provided that the right side is of order $o(r^{*})$. By the discussion following Corollary \ref{cor1}, it is seen that the last condition is satisfied if
$t (n^{-\nu/(2\nu+1)} \vee \sqrt{\log d/n}) \to 0$ is faster than $m^{-1/2} \asymp n^{-1/2(2\nu+1)}$, which is satisfied by the present hypothesis that $t^{2} \{ n^{(1-2\nu)/(2 \nu +1)} \vee (m \log d/n) \} \to 0$. Thus, (\ref{conv}) holds. In view of the proof of Theorem \ref{thm1}, it is not hard to see that 
\begin{equation*}
\| \hat{g} - g_{\bar{\bm{\beta}}} \|_{L_{2}} = \| g_{\hat{\bm{\beta}}} - g_{\bar{\bm{\beta}}} \|_{L_{2}} \lesssim_{p} \phi_{\max} \| \hat{\bm{\beta}} - \bar{\bm{\beta}} \|_{2} \lesssim \| \hat{\bm{\beta}} - \bar{\bm{\beta}} \|_{2} \lesssim_{p} t (n^{-\nu/(2\nu+1)} \vee \sqrt{\log d/n}).
\end{equation*} 
The desired conclusion now follows from the relation:
\begin{align*}
\| \hat{g} - g \|_{L_{2}} \leq \| \hat{g} - g_{\bar{\bm{\beta}}} \|_{L_{2}} + \| g_{\bar{\bm{\beta}}} - g \|_{L_{2}} &\lesssim_{p} t (n^{-\nu/(2\nu+1)} \vee \sqrt{\log d/n}) + m^{-\nu} \\
&\asymp t (n^{-\nu/(2\nu+1)} \vee \sqrt{\log d/n}).
\end{align*}
\end{proof}

\begin{proof}[Proof of Lemma \ref{lem7}]
Invoke that by (\ref{norm})  $\bm{\Sigma} = \diag ( 1,\bm{\Psi},\dots,\bm{\Psi} )$, so that for $\bm{\beta} = (\beta_{G_{1}},\bm{\beta}_{G_{2}}',\dots,\bm{\beta}_{G_{q}}')' \ (q = d+1)$, 
\begin{equation*}
\bm{\beta}'\bm{\Sigma}\bm{\beta} = \beta_{G_{1}}^{2} + \sum_{k=2}^{q} \bm{\beta}_{G_{k}}'\bm{\Psi}\bm{\beta}_{G_{k}}.
\end{equation*}
Thus, by the present hypothesis, there exist positive constants $c$ and $C \ (c < C)$ independent of $n$ such that for all $\bm{\beta} = (\beta_{G_{1}},\bm{\beta}_{G_{2}}',\dots,\bm{\beta}_{G_{q}}')' \in \mathbb{R}^{1+dm}$, 
\begin{equation*}
(1 \wedge c) \| \bm{\beta} \|^{2}_{2} \leq \beta_{G_{1}}^{2} + c \sum_{k=2}^{q} \| \bm{\beta}_{G_{k}} \|^{2}_{2} \leq \bm{\beta}'\bm{\Sigma}\bm{\beta} \leq \beta_{G_{1}}^{2} + C \sum_{k=2}^{q} \| \bm{\beta}_{G_{k}} \|^{2}_{2} \leq (1 \vee C) \| \bm{\beta} \|_{2}^{2}.
\end{equation*}
This implies the desired conclusion.  
\end{proof}

\newpage 

\begin{longtable}{c|ccc|ccc|ccc}
\caption{Simulation results for Model 1}
\label{table1} \\
\hline
& \multicolumn{9}{c}{Case 1} \\
\hline
& \multicolumn{3}{c}{$\tau =0.25$} & \multicolumn{3}{c}{$\tau=0.5$} & \multicolumn{3}{c}{$\tau=0.75$}\\
\hline
 & NSG & NSV & RMSE & NSG & NSV & RMSE & NSG & NSV & RMSE \\
 \hline 
GRLasso & 2.01 & 6.07 & 0.507 & 2.01 & 6.06 & 0.440 & 2.01 & 6.03 & 0.500 \\
& (0.11) & (0.57) & (0.102) & (0.11) & (0.54) & (0.081)  & (0.07) & (0.35) & (0.096) \\
Lasso & 2.02 & 6.01 & 1.028 & 2.02 & 6.02 & 0.803 & 2.02 & 6.01 & 1.029 \\
& (0.15) & (0.18) & (0.233) & (0.15) & (0.15) & (0.152) & (0.12)  & (0.16) & (0.229) \\
QR & 101.00 & 501.00 & 1.984 & 101.00 & 501.00 & 1.936 & 101.00 & 501.00 & 1.985 \\
& (0.00) & (0.08) & (0.057) & (0.00) & (0.05) & (0.056) & (0.00) & (0.05) & (0.056)  \\
\hline
& \multicolumn{9}{c}{Case 2} \\
\hline
GRLasso & 3.44 & 13.21 & 2.251  & 4.72  & 19.60 &  1.929 & 3.36 & 12.77  & 2.261  \\
& (1.13) & (5.66) & (0.162) & (1.05) & (5.27) & (0.184)  & (1.15) & (5.72) & (0.151) \\
Lasso & 5.00 & 5.00 & 1.894 & 5.84  & 5.85  & 1.428 & 4.93 &  4.94 & 1.901  \\
& (1.09) & (1.09) & (0.323) & (0.49) & (0.51) & (0.280) & (1.10)  & (1.10) & (0.314) \\
QR & 100.00 & 501.00 & 2.115 & 100.00  & 500.99 & 2.070 & 100.00 & 501.00 & 2.114 \\
& (0.00) & (0.05) & (0.049) & (0.00) & (0.08) & (0.048) & (0.00) & (0.06) & (0.048)  \\
\hline

\caption*{{\footnotesize ``GRLasso'' refers to  the group Lasso estimator, ``Lasso'' to the $\ell_{1}$-penalized quantile regression estimator, ``QR'' to the unpenalized quantile regression estimator, ``NSG'' to the number of selected groups  (including $G_{1}$), ``NSV''  to the number of selected variables (among $x_{i1},\dots,x_{ip}$), and ``RMSE'' to the root mean squared error $\sqrt{\mathrm{E}[\| \hat{\bm{\beta}} - \bar{\bm{\beta}} \|_{2}^{2}]}$. Standard deviations are given in parentheses. }}
\end{longtable}

\newpage

\begin{longtable}{c|cc|cc|cc}
\caption{Simulation results for Model 2}
\label{table2} \\
\hline
& \multicolumn{6}{c}{$n=400$} \\
\hline 
& \multicolumn{2}{c}{$\tau =0.25$} & \multicolumn{2}{c}{$\tau=0.5$} & \multicolumn{2}{c}{$\tau=0.75$} \\
\hline
 & NSV &  RMSE & NSV & RSME & NSV & RMSE \\
\hline
GRLasso & 3.13 & 0.383 & 3.17 & 0.314 & 3.14 & 0.351 \\
&              (0.35) &  (0.062)   & (0.43) & (0.044) & (0.37) & (0.050) \\ 
Lasso & 3.14 & 0.424 & 3.15 & 0.357 & 3.12 & 0.377 \\
& (0.38) & (0.048) & (0.39) & (0.030) & (0.34) & (0.036) \\
QR & 100.00 &  1.611 & 100.00 & 1.584 & 100.00 & 1.614 \\
& (0.00) & (0.154) & (0.00) & (0.156) & (0.00) & (0.154) \\
\hline
& \multicolumn{6}{c}{$n=800$} \\
\hline
GRLasso & 3.16 & 0.241 & 3.19 & 0.211 & 3.18 & 0.232 \\
&              (0.39) &  (0.032)   & (0.44) & (0.026) & (0.43) & (0.030) \\ 
Lasso & 3.12 & 0.323 & 3.14 & 0.300 & 3.14 &  0.308 \\
& (0.34) & (0.020) & (0.38) & (0.014) & (0.37) & (0.016) \\
QR & 100.00  &  2.224 &  100.00 & 2.100 & 100.00 & 2.229 \\
& (0.00) & (0.197) & (0.00) & (0.182) & (0.00) & (0.191) \\
\hline

\caption*{{\footnotesize ``GRLasso''  refers to  the group Lasso estimator, ``Lasso''  to the $\ell_{1}$-penalized quantile regression estimator, ``QR''  to the unpenalized quantile regression estimator, ``NSV''  to the number of selected variables (among $z_{i1},\dots,z_{i,100}$), and ``RMSE'' refers to the root mean squared error $\sqrt{\mathrm{E}[ \| g_{\hat{\bm{\beta}}} - g \|_{L_{2}}^{2}]}$. Standard deviations are given in parentheses. }}
\end{longtable}

\begin{thebibliography}{99}
\bibitem[Alizadeh and Goldfarb(2003)]{AG03}
Alizadeh, F. and Goldfarb, D. (2003). Second-order cone programming. {\em Math. Program. Ser.B} \textbf{95} 3-51.
\bibitem[Bach(2008)]{B08}
Bach, F.R. (2008). Consistency of the group lasso and multiple kernel learning. {\em J. Mach. Learn. Res.} \textbf{9} 1179-1225.
\bibitem[Belloni and Cernozhukov(2009)]{BC09}
Belloni, A. and Chernozhukov, V. (2009). Post $\ell_{1}$-penalized estimators in high-dimensional linear regression models. Preprint. 
\bibitem[Belloni and Chernozhukov(2011)]{BC10}
Belloni, A. and Chernozhukov, V. (2011). $\ell_{1}$-penalized quantile regression in high-dimensional sparse models. {\em Ann. Statist.} \textbf{39} 82-130.
\bibitem[Bickel et al.(2009)]{BRT09}
Bickel, P., Ritov, Y. and Tsybakov, A. (2009). Simultaneous analysis of Lasso and Dantzig selector. {\em Ann. Statist.} \textbf{37} 1705-1732.
\bibitem[Bousquet(2002)]{B02}
Bousquet, O. (2002). A Bennet concentration inequality and its application to suprema of empirical processes. {\em C.R. Math. Acad. Sci. Paris} \textbf{334} 495-500.
\bibitem[Bunea et al.(2007a)]{BTW07a}
Bunea, F., Tsybakov, A. and Wegkamp, M. (2007). Aggregation for Gaussian regression. {\em Ann. Statist.} \textbf{35} 1674-1697.
\bibitem[Bunea et al.(2007b)]{BTW07b}
Bunea, F., Tsybakov, A. and Wegkamp, M. (2007). Sparsity oracle inequalities for the Lasso. {\em Electron. J. Stat.} \textbf{1} 169-184.
\bibitem[Dudley(1999)]{D99}
Dudley, R.M. (1999). {\em Uniform Central Limit Theorem}. Cambridge University Press.
\bibitem[He and Shao(2000)]{HS00}
He, X. and Shao, Q.-M. (2000). On parameters of increasing dimensions. {\em J. Multivariate Anal.} \textbf{73} 120-135.
\bibitem[Horowitz and Lee(2005)]{HL05}
Horowitz, J.L. and Lee, S. (2005). Nonparametric estimation of an additive quantile regression model. {\em J. Amer. Stat. Assoc.} \textbf{100} 1238-1249.
\bibitem[Huang et al.(2010)]{HHW10}
Huang, J., Horowitz, J.L. and Wei, F. (2010). Variable selection in nonparametric additive models. {\em Ann. Statist.} \textbf{38} 2282-2313.
\bibitem[Huang and Zhang(2010)]{HZ10}
Huang, J. and Zhang, T. (2010). The benefit of group sparsity. {\em Ann. Statist.} \textbf{38} 1978-2004.
\bibitem[Knight(1998)]{K98}
Knight, K. (1998). Limiting distributions for $L_1$ regression estimators under general conditions. {\em Ann. Statist.} \textbf{26} 755-770.
\bibitem[Koenker(2005)]{K05}
Koenker, R. (2005). {\em Quantile Regression}. Oxford University Press.
\bibitem[Koenker and Bassett(1978)]{KB78}
Koenker, R. and Bassett, G. (1978). Regression quantiles. {\em Econometrica} \textbf{46} 33-50.
\bibitem[Koltchinskii and Yuan(2010)]{KY10}
Koltchinskii, V. and Yuan, M. (2010). Sparsity in multiple kernel learning. {\em Ann. Statist.}, to appear.
\bibitem[Lobo et al.(1998)]{LVBL98}
Lobo, M.S., Vandenberghe, L., Boyd, S. and Lebret, H. (1998). Applications of second-order cone progamming. {\em Linear Algebra and its Applications} \textbf{284} 193-228.
\bibitem[Lounici et al.(2010)]{LPTV10}
Lounici, K., Pontil, M., Tsyvakov, A.B. and van de Geer, S.A. (2010). Oracle inequalities and optimal inference under group sparsity. Preprint.  
\bibitem[Ledoux and Talagrand(1991)]{LT91}
Ledoux, M. and Talagrand, M. (1991). {\em Probability in Banach Spaces}. Springer-Verlag.
\bibitem[Meier et al.(2008)]{MVB08}
Meier, L., van de Geer, S.A. and Buhlmann, P. (2008). The group lasso for logistic regression. {\em J.R. Stat. Soc. Ser. B Stat. Methodol.} \textbf{70} 53-71.
\bibitem[Meier et al.(2009)]{MVB09}
Meier, L., van de Geer, S.A. and Buhlmann, P. (2009). High-dimensional additive modeling. {\em Ann. Statist.} 3779-3821.
\textbf{37} 3779-3821.
\bibitem[Meinshausen and Yu(2009)]{MY09}
Meinshausen, N. and Yu, B. (2009). Lasso-type recovery and sparse representations for high-dimensional data. {\em Ann. Statist.} \textbf{37} 246-270.
\bibitem[Mendelson et al.(2007)]{MPT-J07}
Mendelson, S., Pajor, A. and Tomaczak-Jaegermann, N. (2007). Reconstruction and subgaussian operators in asymptotic geometric analysis. {\em Geom. Func. Anal.} \textbf{17} 1248-1282.
\bibitem[Mendelson et al.(2008)]{MPT-J08}
Mendelson, S., Pajor, A. and Tomaczak-Jaegermann, N. (2008). Uniform uncertainty principle for Bernoulli and subgaussian ensembles. {\em Const. Approx.} \textbf{28} 277-289.
\bibitem[Nardi and Rinardo(2008)]{NR08}
Nardi, Y. and Rinardo, A. (2008). On the asymptotic properties of the group lasso estimator for linear models. {\em Electron. J. Stat.} \textbf{2} 605-633.
\bibitem[Nagahban et al.(2010)]{NRWY10}
Negahban, S., Ravikumar, P., Wainwright, M.J. and Yu, B. (2010).  A unified framework for high-dimensional analysis of M-estimators with decomposable regularizers. Preprint. 
\bibitem[Newey(1997)]{N97}
Newey, W.K. (1997). Convergence rates and asymptotic normality for series estimators. {\em J. Econometrics} \textbf{79} 147-168. 
\bibitem[Obozinski et al.(2010)]{OWJ10}
Obozinski, G., Wainwright, M.J. and Jordan, M.I. (2010). Union support recovery in high-dimensional multivariate regression. {\em Ann. Statist.}, to appear.
\bibitem[Oliveira(2010)]{O10}
Oliveira, R.I. (2010). Sums of random Hermitian matrices and an inequality by Rudelson. {\em Elec. Comm. in Probab.} \textbf{15} 203-212. 
\bibitem[Raskutti et al.(2010)]{RWY10}
Raskutti, G., Wainwright, M.J. and Yu, B. (2010). Minmax-optimal rates for sparse additive models over kernel classes via convex programming. Preprint. 
\bibitem[Ravikumar et al.(2009)]{RLLW09}
Ravikumar, P.,  Liu, H., Lafferty, J. and Wasserman, L. (2009). Sparse additive models. {\em J.R. Stat. Soc. Ser. B Stat. Methodol.} \textbf{71} 1009-1030.
\bibitem[Rudelson(1999)]{R99}
Rudelson, M. (1999). Random vectors in the isotropic position. {\em J. Funct. Anal.} \textbf{164} 60-72.
\bibitem[Stone(1982)]{S82}
Stone, C.J. (1982). Optimal global rates of convergence for nonparametric regression. {\em Ann. Statist.} \textbf{10} 1040-1053. 
\bibitem[Stone(1985)]{S85}
Stone, C.J. (1985). Additive regression and other nonparametric models. {\em Ann. Statist.} \textbf{13} 689-705. 
\bibitem[Stout(1974)]{S74}
Stout, W.F. (1974). {\em Almost Sure Convergence}. Academic Press. 
\bibitem[Sturm(1999)]{S99}
Sturm, J. (1999). Using SeDuMi 1.02, a MATLAB toolbox for optimization over symmetric cones. {\em Optim. Methods Software} \textbf{11-12} 625-653. 
\bibitem[Talagrand(1996)]{Ta96}
Talagrand, M. (1996). New concentration inequalities in product spaces. {\em Invent. Math. } \textbf{126}  503-563.
\bibitem[Tibshirani(1996)]{Ti96}
Tibshirani, R. (1996).  Regression shrinkage and selection via the lasso. {\em J.R. Stat. Soc. Ser. B Stat. Methodol.} \textbf{58} 267-288.
\bibitem[Yuan and Lin (2006)]{YL06}
Yuan, M. and Lin, Y. (2006). Model selection and estimation in regression with grouped variables. {\em J.R. Stat. Soc. Ser. B Stat. Methodol.} \textbf{68} 49-67.
\bibitem[van de Geer(2008)]{vdG08}
van de Geer, S.A. (2008). High-dimensional generalized linear models and the Lasso. {\em Ann. Statist.} \textbf{36} 614-645.
\bibitem[van der Vaart and Wellner(1996)]{VW96}
van der Vaart, A.W. and Wellner, J.A. (1996). {\em Weak Convergence and Empirical Processes: With Applications to Statistics}. Springer-Verlag.
\bibitem[Vershynin(2011)]{V11}
Vershynin, R. (2011). Introduction to non-asymptotic analysis of random matrices. In: {\em Compressed Sensing: Theory and Applications}, eds. Yonina Eldar and Gitta Kutyniok. Cambridge University Press, to appear.
\bibitem[Wainwright(2009)]{W09}
Wainwright, M.J. (2009). Sharp thresholds for high-dimensional and noisy sparsity recovery using L1-constrained
quadratic programming (Lasso). {\em IEEE Trans. Inform. Theory} \textbf{55} 2183-2202.
\bibitem[Wei and Zhang(2010)]{WZ10}
Wei, F. and Zhang, J. (2010). Consistent group selection in high-dimensional linear regression. {\em Bernoulli} \textbf{16} 1369-1384.
\bibitem[Zhao and Yu(2007)]{ZY07}
Zhao, P. and Yu, B. (2007). On model selection consistency of Lasso. {\em J. Mach. Learn. Res.}
\textbf{7} 2541-2567.
\bibitem[Zhang(2009)]{Z09}
Zhang, T. (2009). Some sharp performance bounds for least squares regression with L1 penalization. {\em Ann. Statist.} \textbf{37} 2109-2144.
\bibitem[Zhang and Huang(2008)]{ZH08}
Zhang, C.H. and Huang, J. (2008). The sparsity and bias of the lasso selection in high-dimensional linear
regression. {\em Ann. Statist.} \textbf{36} 1567-1594.
\end{thebibliography}
\end{document}